\documentclass[pdftex,dvipdfm,epsf]{iopart}
\eqnobysec
\usepackage{iopams,amsfonts,rotating,graphicx,fancybox,fancyhdr,esint,dsfont}
\usepackage{appendix}

\newcommand{\mkbox}[3]{\hbox{\vrule
      \vbox to  #1{\hrule \vss
                  \hbox to #2{\hss#3\hss}\vss
                  \hrule}\vrule}}

\begin{document}

\title[Non-Hermitean Wishart random matrices (I)]
        {Non-Hermitean Wishart random matrices (I)}

\author{Eugene Kanzieper$^{1,2}$ and Navinder Singh$^{1,3}$}

\address{
    $^1$ Department of Applied Mathematics, H.I.T. -- Holon Institute of
    Technology, Holon 58102,
    Israel\\
    $^2$ Department of Physics of Complex Systems, Weizmann Institute of Science, Rehovot 76100, Israel\\
    $^3$ Chemical Physics Theory Group, Department of Chemistry, University of Toronto, Toronto, Ontario M5S 3H6, Canada}
    \eads{\mailto{eugene.kanzieper@hit.ac.il},
    \mailto{navinder.phy@gmail.com}}

\begin{abstract}
A non-Hermitean extension of paradigmatic Wishart random matrices is introduced to set up a theoretical
framework for statistical analysis of (real, complex and real-quaternion) stochastic time series representing two
`remote' complex systems. The first paper in a series provides a detailed spectral theory of non-Hermitean Wishart random matrices composed of complex
valued entries. The great emphasis is placed on an asymptotic analysis of the mean eigenvalue density for which we derive, among other results, a complex-plane analogue of the Mar\u{c}enko-Pastur law. A surprising connection with a class of matrix models previously invented in the context of quantum chromodynamics is pointed out.
\end{abstract}

\pacs{02.10.Yn, 02.50.-r, 05.40.-a, 75.10.Nr
\\
{}\\
\textsc{Journal of Mathematical Physics \textbf{51}, 103510 (2010)}
}
\newpage
\tableofcontents
\newpage
\section{Introduction} \label{Sec1}

\subsection{Hermitean Wishart random matrices: A brief overview}\label{Sec1-1}
Empirical covariance matrices ${\boldsymbol {\mathcal C}}$ is a central object in statistical analysis of dynamical complex systems characterised by multivariate time series. Built upon $n$ discretised sets of `experimental' signals $\{{\boldsymbol x}_1(t),\dots,{\boldsymbol x}_n(t)\}$ of the length $m$ each \footnote{The $\alpha$-th signal ${\boldsymbol x}_\alpha(t)$ is a vector consisting of $m$ components: ${\boldsymbol x}_\alpha(t) = \{x_\alpha(t_1),\dots, x_\alpha(t_m)\}$.},
\begin{eqnarray}
\label{c-def}
    {\mathcal C}_{\alpha\alpha^\prime} = \sum_{j=1}^m x_\alpha(t_j) \, \bar{x}_{\alpha^\prime}(t_j),\quad (\alpha,\alpha^\prime) \in (1,\dots,n),
\end{eqnarray}
these matrices are {\it intrinsically noisy}. The experimental data $\{{\boldsymbol x}_\alpha(t)\}$ can be as diverse as (i) time-dependent price changes of various assets (Laloux, Cizeau, Bouchaud and Potters 1999; Plerou, Gopikshnan, Rosenow, Amaral and Stanley 1999; Plerou, Gopikshnan, Rosenow, Amaral, Guhr and Stanley 2002) at a stock exchange; (ii) multichannel physiological (e.g., electro- and magnetoencephalography) recordings (Kwapie\'n, Dro\.zd\.z, Liu and Ioannides 1998; Kwapie\'n, Dro\.zd\.z and Ioannides 2000; \v{S}eba 2003); (iii) an atmospheric parameter (such as wind velocity, geopotential height, or temperature) as a function of time and a discretised coordinate (Santhanam and Patra 2001); (iv) information flows in the world-wide web (Barth\'elemy, Gondran and Guichard 2002); (v) pixel decomposed signal of a noisy image (Basu, Ray and Panigrahi 2010), and many more.
\newline\newline\noindent
The random content of empirical covariance matrices originates from (i) finiteness of the time series used (when the length $m$ of a time series is not very large compared to $n$, the number of experimental signals) and (ii) non-stationarity of system correlations (Plerou, Gopikshnan, Rosenow, Amaral and Stanley 2000). As this intrinsic noise blurs the information about genuine correlations in a complex system, empirical covariance matrices must be `cleaned'. A pretty efficient denoising can be achieved by confronting statistics of eigenvalues and eigenvectors of empirical covariance matrices ${\boldsymbol {\mathcal C}}$ with those of {\it most random covariance matrices} ${\boldsymbol {\mathcal W}}$ constructed from mutually uncorrelated (Gaussian) times series. Evidently, the deviations in statistics of ${\boldsymbol {\mathcal C}}$ and of ${\boldsymbol {\mathcal W}}$ are to feature the properties that are specific to a complex system and can therefore be used to extract genuine correlations between various system components out of experimental data ${\boldsymbol x}_\alpha(t)$. In this paper, we shall only be concerned with the {\it spectral analysis} of covariance matrices. For an introductory exposition of the eigenvector-based analysis the reader is referred to the review by Bouchaud and Potters (2009).
\newline\newline\noindent
The most random covariance matrices, also known as the Wishart matrices (Wishart 1928), are fundamental to the whole realm of multivariate statistical analysis (Muirhead 1982). Given $n$ sets of uncorrelated discretised zero-mean Gaussian random processes $X_\alpha(t)$, an $n\times n$ random Wishart matrix ${\boldsymbol {\mathcal W}}$ is defined by its entries
\begin{eqnarray}
    {\mathcal W}_{\alpha\alpha^\prime} = \sum_{j=1}^m X_\alpha(t_j) \bar{X}_{\alpha^\prime}(t_j),\quad (\alpha,\alpha^\prime) \in (1,\dots,n).
\end{eqnarray}
In a matrix notation,
\begin{eqnarray}
\label{eq-12}
    {\boldsymbol {\mathcal W}} = {\boldsymbol {\mathcal X}} {\boldsymbol {\mathcal X}}^\dagger,
\end{eqnarray}
where ${\boldsymbol {\mathcal X}}$ denotes a rectangular $n\times m$ matrix whose $(\alpha,j)$ entry equals ${\mathcal X}_{\alpha,j}=X_\alpha(t_j)$, and $m$ is the length of a time series.  The associated probability measure reads
\begin{eqnarray}\label{eq-12a}
    dP_{n,m}^{(\beta)}({\boldsymbol {\mathcal X}}) = c_{n,m}^{-1}(\beta, a_\beta)\, \exp\left[
      - a_\beta \,{\rm tr\,} {\boldsymbol {\mathcal X}}{\boldsymbol {\mathcal X}}^\dagger
    \right]\, D{\boldsymbol {\mathcal X}}.
\end{eqnarray}
Here, $c_{n,m}(\beta)$ is the normalisation constant,
\begin{eqnarray}\label{eq-12b}
    c_{n,m}(\beta, a_\beta) =
    \left(
        \frac{\pi}{a_\beta}
    \right)^{\beta n m/2},
\end{eqnarray}
the notation $D{\boldsymbol {\mathcal X}}$ stands for
\begin{eqnarray}\label{eq-12c}
    D{\boldsymbol {\mathcal X}} = \prod_{\alpha=1}^n \prod_{j=1}^m \left(
            \prod_{s=1}^\beta d{\mathcal X}_{\alpha,j}^{(s)}
            \right),
\end{eqnarray}
and $\beta$ denotes the Dyson index (Mehta 2004). The superscript $(s)$ in Eq.~(\ref{eq-12c}) counts a number of degrees of freedom in $X_\alpha(t_j)$. For real, complex and quaternion real time series, $\beta$ equals $1$, $2$, and $4$, respectively. Equations~(\ref{eq-12}) -- (\ref{eq-12c}) uniquely define three canonical ensembles of random Wishart matrices.
\newline\newline\noindent
By definition, the canonical Wishart probability measure $d\pi_{n,m}^{(\beta)}({\boldsymbol {\mathcal W}})$ is provided by the matrix integral
\begin{eqnarray}
 \label{eq-16a}
    d\pi_{n,m}^{(\beta)}({\boldsymbol {\mathcal W}}) =
    \int dP_{n,m}^{(\beta)}({\boldsymbol {\mathcal X}})\,
    \delta\left(
            {\boldsymbol {\mathcal W}} - {\boldsymbol {\mathcal X}}{\boldsymbol {\mathcal X}}^\dagger
        \right)\, D{\boldsymbol {\mathcal W}},
\end{eqnarray}
where the flat measure $D{\boldsymbol {\mathcal W}}$ equals
\begin{eqnarray}\label{dw-flat}
D{\boldsymbol {\mathcal W}} = \prod_{\alpha=1}^n d{\mathcal W}_{\alpha\alpha}\prod_{\alpha>\alpha^\prime=1}^n \left(
            \prod_{s=1}^\beta d{\mathcal W}_{\alpha,\alpha^\prime}^{(s)}
            \right).
 \end{eqnarray}
In the Wishart domain ($m \ge n$), this matrix integral can be evaluated in a closed form resulting in the elegant formula (Muirhead 1982) \footnote{The anti-Wishart domain ($m<n$) has been treated by Janik and Nowak (2003).}:
 \begin{eqnarray}
 \label{eq-16b} \fl \quad
    \phantom{\frac{0^0}{0_0}} d\pi_{n,m}^{(\beta)}({\boldsymbol {\mathcal W}}) =\tilde{c}_{n,m}^{-1}(\beta,a_\beta)\,\Theta({\boldsymbol {\mathcal W}})\,\left({\rm det\,}{\boldsymbol {\mathcal W}}\right)^{\beta(n-m+1)/2 -1}
            \, \exp\left[
        - a_\beta \,{\rm tr\,} {\boldsymbol {\mathcal W}}
    \right] \,
    D{\boldsymbol {\mathcal W}}.
\end{eqnarray}
Here, $\tilde{c}_{n,m}(\beta,a_\beta)$ is yet another normalisation constant,
\begin{eqnarray} 
    \tilde{c}_{n,m}(\beta,a_\beta) = \frac{\pi^{n(n-1)\beta/4}}{a_\beta^{n m \beta/2} } \prod_{j=1}^n \Gamma\left(
        \frac{\beta}{2}(m+j-1)
    \right).
\end{eqnarray}
The Heaviside function $\Theta({\boldsymbol {\mathcal W}})$ indicates that ${\boldsymbol {\mathcal W}}$ is a positive definite matrix.
\newline\newline\noindent
Spectral statistical properties of the Wishart random matrices are well understood since the probability measure in the eigenvalue space
readily follows from Eq.~(\ref{eq-16b}) upon diagonalisation of ${\boldsymbol {\mathcal W}}$ (Mehta 2004):
\begin{eqnarray}\fl
\label{wish-jpdf}
    d\rho_{n,m}^{(\beta)}(\lambda_1,\dots,\lambda_n)
    = \hat{c}_{n,m}^{-1}(\beta,a_\beta)\,\prod_{j=1}^n \Theta(\lambda_j) \,\lambda_j^{\beta(m-n+1)/2 -1} e^{-a_\beta\lambda_j} |\Delta_n({\boldsymbol \lambda})|^\beta
    \prod_{j=1}^n d\lambda_j.\nonumber\\{}
\end{eqnarray}
Here, $\hat{c}_{n,m}(\beta,a_\beta)$ is the normalisation constant,
\begin{eqnarray} \fl
    \hat{c}_{n,m}(\beta,a_\beta) = \frac{1}{a_\beta^{n m \beta/2} \left[\Gamma(1+\beta/2)\right]^n} \prod_{j=1}^n \Gamma\left(
        1+ \frac{\beta}{2}\,j
    \right) \, \Gamma\left(
        1+ \frac{\beta}{2}\,(m-n+j)
    \right).
\end{eqnarray}
Spectral correlation functions of arbitrary finite order can be restored from Eq.~(\ref{wish-jpdf}) by applying the orthogonal polynomial technique (Mehta 2004). The mean density of eigenlevels $\varrho_{n,m}^{(\beta)}(\lambda)$ is of primary importance for denoising of empirical covariance matrices. Its large-$n$ limit (when the ratio $q=m/n\ge 1$ is kept fixed) is described by the Mar\u{c}enko--Pastur law (Mar\u{c}enko and Pastur 1967) \footnote{The Dyson fluid approach is presumably the shortest way to derive this result, see Dyson (1971).}:
\begin{eqnarray}
\label{mp-1}
    \varrho_{n,m}^{(\beta)}(\lambda) = \frac{a_\beta}{\beta \pi}\cdot \frac{1}{\lambda} \sqrt{(\lambda_{\rm max} -\lambda)(\lambda-\lambda_{\rm min})},
\end{eqnarray}
where $\lambda \in (\lambda_{\rm min},\lambda_{\rm max})$. The endpoints of the eigenvalue support are
\begin{eqnarray}
\label{mp-2}
     \lambda_{\left\{{\rm min} \atop {\rm max}\right\}} = \frac{n \beta}{2a_\beta}\left(\sqrt{q} \mp 1\right)^2.
\end{eqnarray}
Equations (\ref{mp-1}) and (\ref{mp-2}) lay the basis for a cleaning procedure required to separate information-carrying correlations from the useless noise which is always present in empirical covariance matrices (Bouchaud and Potters 2009).

\subsection{Non-Hermitean Wishart random matrices}
\subsubsection{Motivation and definitions}\label{Sec-1-2-1}
\noindent\newline\newline
Empirical covariance matrices ${\boldsymbol {\mathcal C}}$ discussed in the previous Section have a built-in Hermiticity [Eq.~(\ref{c-def})]. However, in many applied problems of statistical analysis the empirical covariance matrix can equally be {\it non-Hermitean}. This is the case when one is interested in studying correlations between two `remote' complex systems characterised by two distinct sets of time series $\{{\boldsymbol x}_1(t),\dots, {\boldsymbol x}_n(t)\}$ and $\{{\boldsymbol y}_1(t),\dots, {\boldsymbol y}_n(t)\}$, respectively. These series may represent multichannel magneto-encephalography recordings of activity in the left and right auditory cortex (Kwapie\'{n}, Dro\.{z}d\.{z} and Ioannides 2000) or describe return of stocks traded on two large but geographically distant markets (Kwapie\'{n}, Dro\.{z}d\.{z}, G\'{o}rski and O\'{s}wi\c{e}cimka 2006). In both cases, a noise-dressed empirical covariance matrix
\begin{eqnarray}
\label{emp-cov}
    \tilde{\mathcal C}_{\alpha\alpha^\prime} = \sum_{j=1}^m x_\alpha(t_j)\, \bar{y}_{\alpha^\prime}(t_j),\quad (\alpha,\alpha^\prime) \in (1,\dots,n),
\end{eqnarray}
is manifestly non-Hermitean. Consequently, its spectrum as well as the spectrum of an appropriate most random covariance matrix
\begin{eqnarray}
\label{nhw-def}
    {\tilde {\mathcal W}}_{\alpha\alpha^\prime} = \sum_{j=1}^m X_\alpha (t_j) \, {\bar Y}_{\alpha^\prime} (t_j),\qquad (\alpha,\alpha^\prime) \in (1,\dots,n),
\end{eqnarray}
composed of $2n$ sets of uncorrelated discretised zero-mean Gaussian random processes ${\boldsymbol X}_\alpha (t)=\{X_\alpha(t_1),\dots, X_\alpha(t_m)\}$ and ${\boldsymbol Y}_\alpha(t)=\{ Y_\alpha(t_1),\dots, Y_\alpha(t_m)\}$, becomes complex valued. The latter observation has been beautifully illustrated in the studies by Kwapie\'{n}, Dro\.{z}d\.{z} and Ioannides (2000) and Kwapie\'{n}, Dro\.{z}d\.{z}, G\'{o}rski and O\'{s}wi\c{e}cimka (2006).
\newline\newline\noindent
This reasoning leads us to introduce three {\it non-Hermitean} counterparts of canonical Wishart matrix models [Eqs.~(\ref{eq-12}) and (\ref{eq-12a})] defined as
\begin{eqnarray}
\label{eq-12nh}
    \tilde{{\boldsymbol {\mathcal W}}} = {\boldsymbol {\mathcal X}} {\boldsymbol {\mathcal Y}}^\dagger.
\end{eqnarray}
The probability measures assigned to the matrices ${\boldsymbol {\mathcal X}}$ and ${\boldsymbol {\mathcal Y}}$ read
\begin{eqnarray}\label{eq-12a-nhw}
    dP_{n,m}^{(\beta)}({\boldsymbol {\mathcal X}}) = c_{n,m}^{-1}(\beta,a_\beta)\, \exp\left[
      - a_\beta \,{\rm tr\,} {\boldsymbol {\mathcal X}}{\boldsymbol {\mathcal X}}^\dagger
    \right]\, D{\boldsymbol {\mathcal X}}, \\
    \label{eq-12a-nhw-2}
    dP_{n,m}^{(\beta)}({\boldsymbol {\mathcal Y}}) = c_{n,m}^{-1}(\beta,a_\beta^\prime)\, \exp\left[
      - a_\beta^\prime \,{\rm tr\,} {\boldsymbol {\mathcal Y}}{\boldsymbol {\mathcal Y}}^\dagger
    \right]\, D{\boldsymbol {\mathcal Y}}.
\end{eqnarray}
The normalisation constant $c_{n,m}(\beta,a_\beta)$ and the flat measures $D{\boldsymbol {\mathcal X}}$ and $D{\boldsymbol {\mathcal Y}}$ are specified by Eqs.~(\ref{eq-12b}) and (\ref{eq-12c}), respectively.
\newline\newline\noindent
Given the above definition of non-Hermitean Wishart random matrices, we would like to determine (i) the probability measure induced on $\tilde{{\boldsymbol {\mathcal W}}}$, (ii) the joint probability density function of all $n$ complex eigenvalues of $\tilde{{\boldsymbol {\mathcal W}}}$, (iii) the eigenvalue correlation functions of arbitrary (finite) order, and also analyse (iv) the mean spectral density in various scaling limits. Even though the real valued time series ($\beta=1$) are of most interest for statistical applications, the present paper will focus on a somewhat simpler case of complex valued time series ($\beta=2$) \footnote{A detailed spectral theory of non-Hermitean Wishart random matrices at $\beta=1$ will be a subject of a separate publication.}.

\subsubsection{Main results and discussion}\label{Sec-1-2-2}\noindent\newline\newline
For the benefit of the readers, we collect our main results into this easy to read subsection with pointers to the sections containing detailed derivation of each statement.
\newline\newline\noindent
{\it (i)~The probability measure in the matrix space.}---Let $\tilde{\boldsymbol {\mathcal W}}$ be a matrix drawn from an ensemble of $n\times n$ complex non-Hermitean Wishart random matrices as defined by Eqs.~(\ref{eq-12nh}), (\ref{eq-12a-nhw}) and (\ref{eq-12a-nhw-2}). Then, for $m\ge n$, the probability measure $d\pi_{n,m}(\tilde{{\boldsymbol {\mathcal W}}})$ associated with the matrix entries $\{{\tilde{\mathcal W}}_{\alpha,\alpha^\prime}\}$ equals
\begin{eqnarray}
\label{mrs-01a}\fl
    {\rm (Type\;I)}\qquad d\pi_{n,m}(\tilde{{\boldsymbol {\mathcal W}}}) = \frac{1}{(2\pi)^{2n^2}}
    \int_{{\mathbb C}^{n\times n}} D{\boldsymbol q}\,
    \exp\left[
         \frac{i}{2}\, {\rm tr} \left(
            {\boldsymbol q} {\tilde{\boldsymbol {\mathcal W}}}^\dagger + {\boldsymbol q}^\dagger {\tilde {\boldsymbol {\mathcal W}}}
        \right)
    \right] \nonumber\\
    \fl \qquad \qquad \qquad \qquad\quad \times
    {\rm det}^{-m} \left(
    {\mathds 1}_n + \frac{1}{4a_2 a_2^\prime} {\boldsymbol q}{\boldsymbol q}^\dagger
    \right)\, D\tilde{{\boldsymbol {\mathcal W}}},
\end{eqnarray}
where the flat measure $D\tilde{{\boldsymbol {\mathcal W}}}$ is defined by
\begin{eqnarray}\label{dw-flat-nh}
    D\tilde{{\boldsymbol {\mathcal W}}} = \prod_{\alpha,\alpha^\prime=1}^n d\mathfrak{Re\,{\tilde{\mathcal W}}_{\alpha,\alpha^\prime}}
    \, d\mathfrak{Im\,{\tilde{\mathcal W}}_{\alpha,\alpha^\prime}}.
\end{eqnarray}
In Eq.~(\ref{mrs-01a}), that will be referred to as the type I representation of $d\pi_{n,m}(\tilde{{\boldsymbol {\mathcal W}}})$, the integration runs over an $n\times n$ generic complex valued matrix.\newline\newline\noindent
An alternative, type II representation,
\begin{eqnarray}\fl \label{mrs-01b}
    {\rm (Type\;II)}\qquad  d\pi_{n,m}(\tilde{{\boldsymbol {\mathcal W}}}) =
    \frac{(a_2 a_2^\prime)^{n(n+m)/2}}{\pi^{n(3n-1)/2}} \frac{\prod_{j=1}^{m-n} \Gamma(j)}{\prod_{j=1}^{m} \Gamma(j)}
    \, \int_{{\boldsymbol Q}^\dagger ={\boldsymbol Q}} D{\boldsymbol Q}\, \Theta({\boldsymbol Q})\,
    \nonumber\\
    \fl \qquad \qquad \qquad \qquad\quad \times
     {\rm det}^{m-2n} {\boldsymbol Q}\,
     \exp\left[-\sqrt{a_2 a_2^\prime} \,{\rm tr\,}( {\boldsymbol Q} + \tilde{{\boldsymbol {\mathcal W}}}^\dagger {\boldsymbol Q}^{-1} \tilde{{\boldsymbol {\mathcal W}}})
    \right]\, D\tilde{{\boldsymbol {\mathcal W}}}.\nonumber\\
    {}
\end{eqnarray}
involves an integral over an $n\times n$ complex Hermitean matrix. Both type I and type II matrix integrals are derived in Section \ref{Sec-2-1} and Appendix \ref{app-2}.
\newline\newline\noindent
{\it (ii)~The joint probability density function of all complex eigenvalues.}---Let $(w_1,\dots,w_n)$ be complex eigenvalues of an $n\times n$ complex non-Hermitean Wishart random matrix.  Their joint probability density function is
\begin{eqnarray} \fl \label{msr-02}
    \rho_{n,m}(w_1,\dots,w_n) = \frac{2^n (a_2 a_2^\prime)^{n(m+1)/2}}{\pi^n n!} \frac{\prod_{j=1}^{m-n}\Gamma(j)}{\prod_{j=1}^n \Gamma(j) \prod_{j=1}^m \Gamma(j)}
    \nonumber\\
    \fl \qquad \qquad \qquad\qquad \times
    |\Delta_n({\boldsymbol w})|^2 \, \prod_{j=1}^n |w_j|^{m-n} \, K_{m-n}\left(
    2 |w_j| \sqrt{a_2 a_2^\prime}
    \right).
\end{eqnarray}
\begin{figure}[t]
\hspace{2cm}
\includegraphics[scale=0.25]{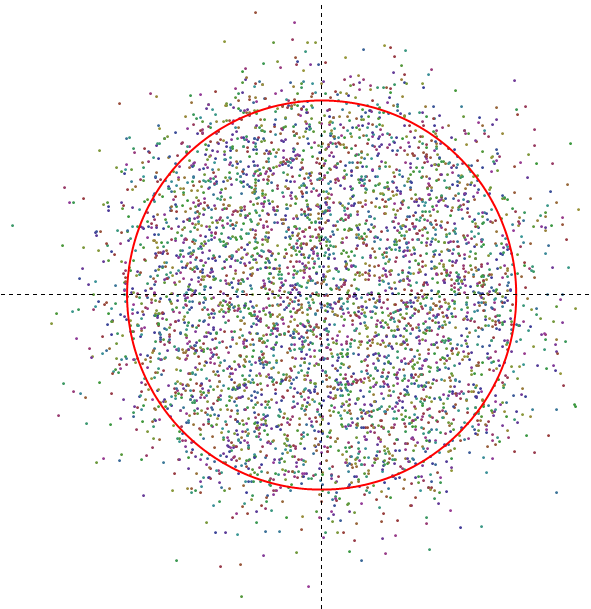}
\quad
\includegraphics[scale=0.25]{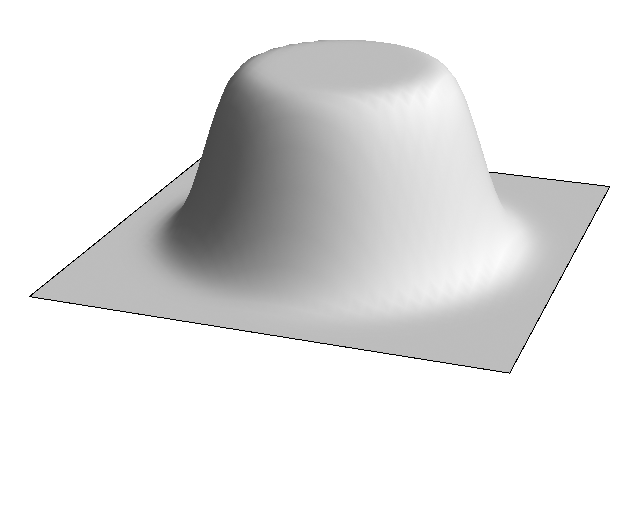}
\caption{Left panel: Numerically simulated distribution of complex eigenvalues in the regime $\{n\; {\rm fixed\; and\;} \nu \gg 1\}$. The red circle of the radius $R=2\sqrt{n\nu}$ is displayed to highlight a notable finite-$n$ dispersion effect. Right panel: Profile of the mean density Eq.~(\ref{eq-ginue-sim}). Both figures correspond to $n=10$ and $\nu = 95$. The simulation shown on the left panel involved diagonalisation of $500$ matrices.}\label{dos-regime-01}
\end{figure}
Here, $K_\nu(w)$ is the modified Bessel function of the second kind. For a detailed derivation, the reader is referred to Section \ref{Sec-2-2-1} and Appendix \ref{app-2}. Interestingly, the above result is a particular case of a more general matrix model introduced by Osborn (2004) in the context of quantum chromodynamics (QCD) with a baryon chemical potential [see also Akemann, Bloch, Shifrin and Wettig (2008)]. In this connection, let us stress that an overlap between our work and the QCD studies cited throughout the paper occurs for the finite-$n$/finite-$\nu$ regime. The large-$n$/large-$\nu$ asymptotic analysis, central to statistical applications which triggered our research, is new, and the techniques used are different from those employed in the QCD literature.
\newline\newline\noindent
{\it (iii)~Determinant structure of spectral correlators.}---The $p$-point correlation function follows from Eq.~(\ref{msr-02}) by virtue of the Dyson integration theorem (Mehta 1976):
\begin{eqnarray}
\label{mrs-03a}
    R^{(p)}_{n,m}(w_1,\cdots,w_p) =
    \, {\rm det} \left[
        {\mathbb K}_{n,m}(w_j, \bar{w}_k)
    \right]_{1\le j,k \le p},
\end{eqnarray}
where \footnote[1]{To simplify notation, we have set $a_2 a_2^\prime=1/4$.}
\begin{eqnarray}\fl \label{mrs-03b}
    {\mathbb K}_{n,m}(w,w^\prime)= \frac{1}{2^{\nu+1} \pi} \left|w w^\prime \right|^{\nu/2} \,\sqrt{K_{\nu}(|w|) K_{\nu}(|w^\prime|)}
    \sum_{k=0}^{n-1} \frac{(w w^\prime)^k} {2^{2k} k!\, (k+\nu)!}\nonumber\\
    {}
\end{eqnarray}
is the two-point scalar kernel. Here and everywhere below
\begin{eqnarray}\label{m-nu}
    \nu = m-n \ge 0.
\end{eqnarray}
See Section \ref{Sec-2-2-2} for a straightforward derivation \footnote{In view of our previous remark in (ii), Eqs.~(\ref{mrs-03a}) and (\ref{mrs-03b}) are automatically consistent with those reported in the QCD literature.}
\newline\newline\noindent
{\it (iv)~The mean density of complex eigenvalues.}---An asymptotic behaviour of the mean spectral density
\begin{eqnarray}
    R_{n,m}^{(1)}(w) = {\mathbb K}_{n,m}(w,\bar{w})
\end{eqnarray}
is of most interest for statistical applications. Depending on the relation between $n$ and $\nu$ (or, equivalently, $m$, see Eq.~(\ref{m-nu})), the following three limiting laws will be established in Section~\ref{Sec-2-3}.
\begin{figure}[t]
\centering
\hspace{2.5cm}
\includegraphics[scale=0.20]{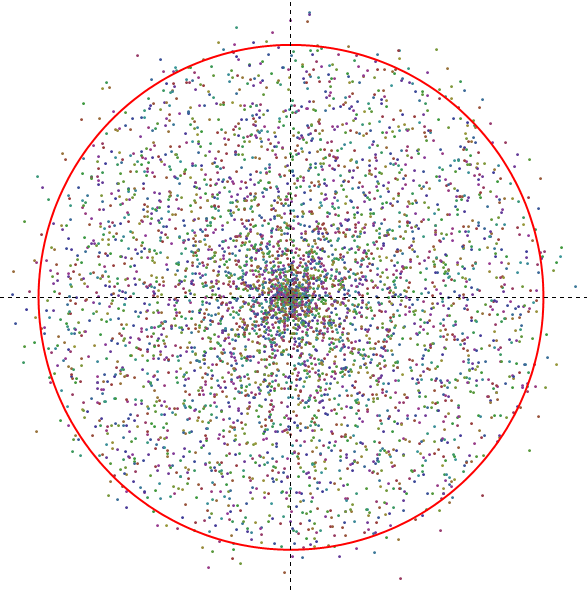}
\quad
\includegraphics[scale=0.25]{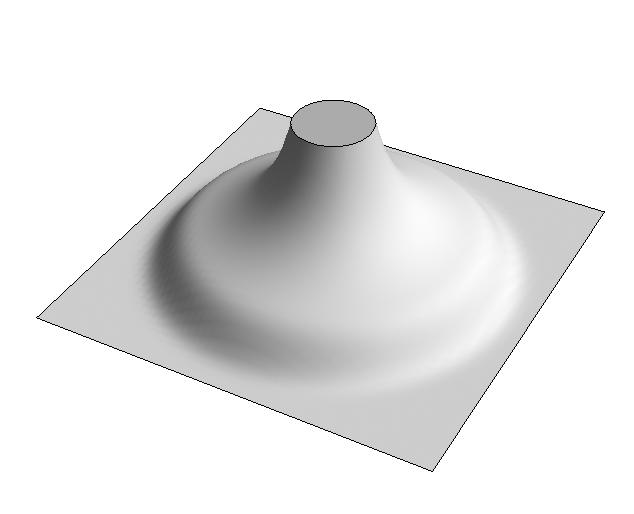}
\caption{Left panel: Numerically simulated distribution of complex eigenvalues in the regime $\{n\gg 1 \; {\rm and\;} \nu {\;\rm{fixed}}\}$. The red circle of the radius $R=2n$ is displayed to indicate the `mean field' edge of the eigenvalue support (see discussion in Section \ref{Sec-2-3-1}). The phenomenon of `clustering' of eigenvalues around the origin stemming from the $1/|w|$ decay Eq.~(\ref{algebraic}) is clearly seen. Right panel: Profile of the mean density Eq.~(\ref{d-global}). A `goblet base' originates from the ${\rm erfc}$-law Eq.~(\ref{erfc-law-part}). Both figures correspond to $n=100$ and $\nu = 1$. The simulation shown on the left panel involved diagonalisation of $100$ matrices.}\label{dos-regime-02}
\end{figure}
\newline\noindent
\begin{itemize}
  \item {\bf Regime I.}---For $n$ fixed and $\nu=m-n \gg 1$, we derive the complementary $\Gamma$-function law
        \begin{eqnarray} \label{eq-ginue-sim}
        R^{(1)}_{n,m}(w) \simeq \frac{1}{4\pi \nu\, \Gamma(n)} \,\Gamma\left(
    n, \displaystyle{\frac{|w|^2}{4\nu}}
    \right).
        \end{eqnarray}
        This formula assumes that the ratio $w/\sqrt{\nu}\sim {\mathcal O}(\nu^0)$ is kept fixed. This result mirrors a finite-$n$ formula for the mean spectral density in the Ginibre unitary ensemble (GinUE) [see, e.g., Eq.~(1.43) in Ginibre (1965) and Eq.~(2.17) in Akemann and Kanzieper (2007)]: the two formulae can be reduced to each other upon a proper rescaling of the energy variable $|w|$.\newline\newline\noindent
        This regime is characterised by (i) a nearly uniform eigenvalue distribution within the disk of the radius $R\simeq 2\sqrt{n\nu}$ and (ii) a notable (finite-$n$) dispersion effect that manifests itself in a significant number of eigenvalues outside the disk. For an illustration, see Fig. \ref{dos-regime-01}.
        \newline
  \item {\bf Regime II.}---For $n \gg 1$ and $\nu=m-n\ge 0$ fixed such that $\nu\sim {\mathcal O}(n^0)$, a complicated behaviour of the mean eigenvalue density is accounted for by the single formula
  \begin{eqnarray}\label{d-global}
        R_{n,m}^{(1)}(w) \simeq \frac{1}{4\pi}\, I_\nu(|w|) \, K_\nu(|w|) \, {\rm erfc\,}\left(\frac{|w|-2n}{2\sqrt{n}}\right).
  \end{eqnarray}
  It incorporates three different regimes.\newline\noindent
  \begin{description}
    \item[{}] (i) If $|w|\sim {\mathcal O}(n^0)$, the above equation reduces to
    \begin{eqnarray}
    \label{eq-d-2}
        R_{\infty}^{(1)}(w) = \frac{1}{2\pi}\, I_\nu(|w|) \, K_\nu(|w|).\\
        {}\nonumber
  \end{eqnarray}
For $\nu=0$, the density exhibits a weak logarithmic singularity around the origin; for $\nu>0$ it stays finite:
\begin{eqnarray}
\label{r-zero-results}
    R_{\infty}^{(1)}(w)\Big|_{|w|\ll 1} = \left\{
               \begin{array}{ll}
                 \displaystyle{\frac{1}{2\pi} \log \frac{1}{|w|}}, & \hbox{$\nu=0$,} \\
                    {} & {}\\
                 \displaystyle{\frac{1}{4\pi \nu}}, & \hbox{$\nu \neq 0$.}
               \end{array}
             \right.
\end{eqnarray}
\begin{figure}[t]
  \centerline{
   \mbox{\includegraphics[scale=0.20]{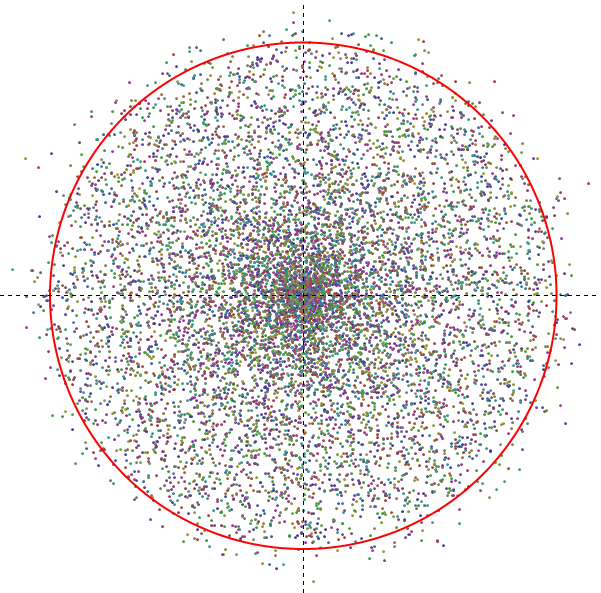}}
   \mbox{\includegraphics[scale=0.20]{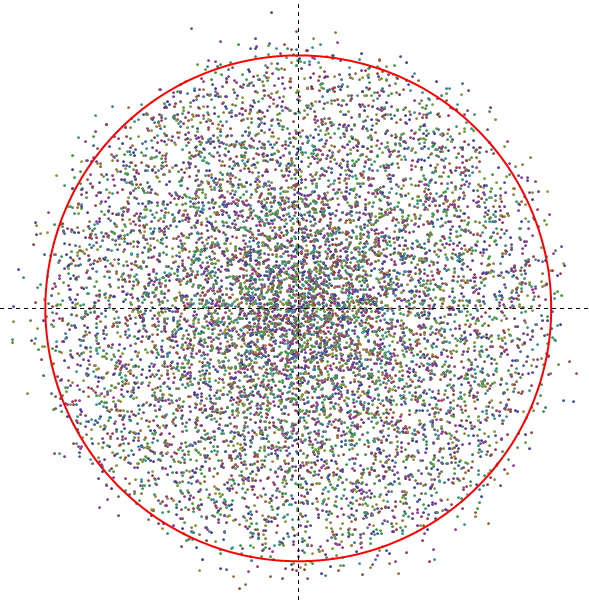}}
   \mbox{\includegraphics[scale=0.20]{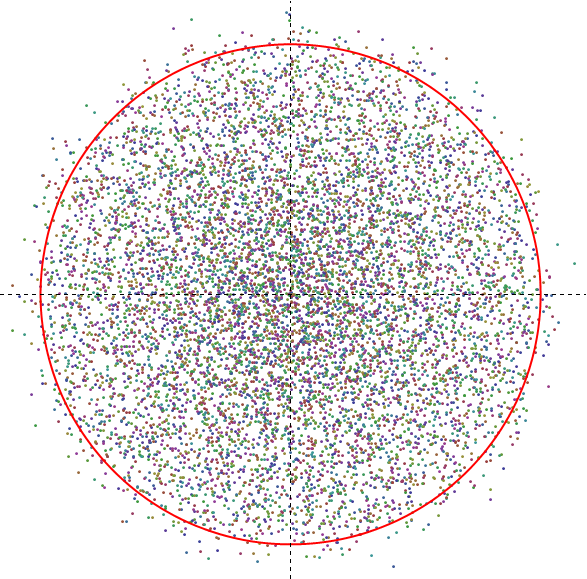}}
}
  \caption{Numerically simulated distribution of complex eigenvalues in the regime $\{n\gg 1$ and $\nu \gg 1$ with $\nu/n=q\;\rm{kept\;fixed}\}$. The red circle of the radius $R=2n\sqrt{1+q}$ is displayed to indicate the `mean field' edge of the eigenvalue support (see discussion in Section \ref{Sec-2-3-3}). The phenomenon of `clustering' of eigenvalues around the origin blurs as $q$ increases. Parameters: $q=1/10$ (left panel), $q=1/2$ (middle panel) and $q=1$ (right panel). The simulation was performed by diagonalising $100$ matrices of the size $100\times 100$.}
  \label{overView}
  \end{figure}
\newline\noindent
    \item[{}] (ii) Increasing $|w|$, one observes that the density Eq.~(\ref{eq-d-2}) exhibits an algebraic decay
  \begin{eqnarray}\label{algebraic}
            R_{\infty}^{(1)}(w) \simeq \frac{1}{4\pi |w|}
  \end{eqnarray}
  that holds for $|w|\gg 1$. This unusual behavior showing up as a `clustering' of complex eigenvalues around the origin (see Fig. \ref{dos-regime-02}), contrasts with the uniform mean density of complex eigenvalues in GinUE that emerges at $|w|\gg 1$ (Ginibre 1965). The `mean field' result Eq.~(\ref{algebraic}) is consistent with a more general observation due to Burda, Janik, and Waclaw (2010) who have advocated existence of the universal $|w|^{-2(1-1/M)}$--law for the mean spectral density of the product of $M$ independent Gaussian random matrices. We also notice that the clustering phenomenon was recently observed in spectra of time-lagged correlation matrices (Biely and Thurner 2008).
  \newline\noindent

    \item[{}] (iii) The $1/|w|$ law breaks down in the $\sqrt{n}$-vicinity of the `critical' point $|w|_c = 2n$, where Eq.~(\ref{d-global}) reduces to
  \begin{eqnarray}\label{erfc-law-part}
            R_{n,m}^{(1)}(w) \simeq \frac{1}{8\pi |w|}\, {\rm erfc\,}\left(\frac{|w|-2n}{2\sqrt{n}}\right).
  \end{eqnarray}
  The complementary error function law describes the tails of the two-dimensional eigenvalue support in the region $|w| - 2n \sim {\mathcal O}(\sqrt{n})$. The ${\rm erfc}$-law can be easily identified as a `goblet base' in the mean density profile shown in Fig. \ref{dos-regime-02} (right panel).  \newline
  \end{description}
  \noindent
  \item {\bf Regime III.}---For $n\gg 1$ and $\nu \gg 1$ such that the ratio $\nu/n=q$ is kept fixed ($q>0$), we prove that the mean density of complex eigenvalues is described by the formula
\begin{eqnarray}
\label{mp-law-cp-2-list}
    R_{n,m}^{(1)}(w) = \frac{1}{8\pi}\,\frac{1}{\sqrt{|w|^2+ n^2 q^2}}\, {\rm erfc}\left(
          \frac{|w|-2n \sqrt{q+1}}{\sqrt{2n(q+2)}}   \right),
\end{eqnarray}
which is an analogue of the Mar\u{c}enko-Pastur law for complex valued eigenvalues. Far away from the critical point $|w_c|=2n\sqrt{q+1}$, the above equation simplifies to
\begin{eqnarray}
\label{mp-law-cp-2-list-red}
    R_{\infty}^{(1)}(w) = \frac{1}{4\pi}\,\frac{1}{\sqrt{|w|^2 + \nu^2}}.\\
    {}\nonumber
\end{eqnarray}
Notice that (formally performed) $\nu \rightarrow 0$ limit reduces Eq.~(\ref{mp-law-cp-2-list}) to Eq.~(\ref{erfc-law-part}) as expected. Figures \ref{overView} and \ref{destruction} show that the phenomenon of clustering of complex eigenvalues around the origin becomes less pronounced with increase of the parameter $q$.
 \end{itemize}
Having announced the main results of our study, we now turn to their detailed derivation.
  \begin{figure}[t]
  \centerline{
  \mbox{\includegraphics[scale=0.28]{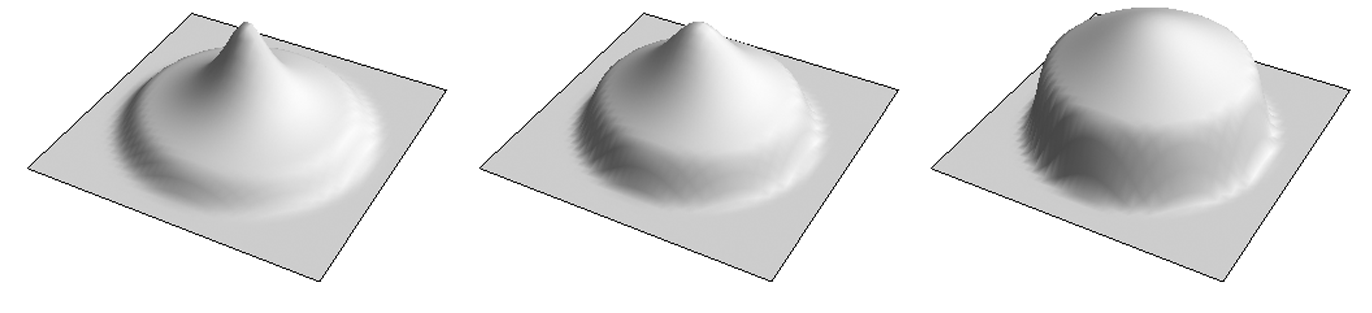}}
  }
  \caption{Destruction of the `clustering' phenomenon in the regime $\{n\gg 1$ and $\nu \gg 1$ with $\nu/n=q\;\rm{kept\;fixed}\}$. Profiles of the mean density Eq.~(\ref{algebraic}) are plotted for $n=100$ and $q=1/2$ (left panel), $q=1$ (middle panel) and $q=3$ (right panel).}
  \label{destruction}
  \end{figure}

\section{Non-Hermitean Wishart random matrices at $\beta=2$}
\subsection{Matrix integral representation of the probability measure}\label{Sec-2-1}
By definition, the probability measure \footnote{From now on, the notation $d\pi_{n,m}(\tilde{{\boldsymbol {\mathcal W}}})$ is kept for the probability measure associated with non-Hermitean Wishart matrices.} $d\pi_{n,m}(\tilde{{\boldsymbol {\mathcal W}}})$ induced on $\tilde{{\boldsymbol {\mathcal W}}}$ is provided by a two-matrix integral
\begin{eqnarray}
\label{m-01}
    d\pi_{n,m}(\tilde{{\boldsymbol {\mathcal W}}}) =
    \int dP_{n,m}^{(2)}({\boldsymbol {\mathcal X}})\,
    \int dP_{n,m}^{(2)}({\boldsymbol {\mathcal Y}}) \, \delta \left(
        \tilde{{\boldsymbol {\mathcal W}}} - {\boldsymbol {\mathcal X}} {\boldsymbol {\mathcal Y}}^\dagger
    \right)\, D\tilde{{\boldsymbol {\mathcal W}}},
\end{eqnarray}
where the flat measure $D\tilde{{\boldsymbol {\mathcal W}}}$ is defined in Eq.~(\ref{dw-flat-nh}). The integration in Eq.~(\ref{m-01}) can be performed in two different ways leading to two different (albeit equivalent) representations of $d\pi_{n,m}(\tilde{{\boldsymbol {\mathcal W}}})$.

\subsubsection{Integral over complex matrix (Type I)}\noindent\newline\newline
We start the evaluation of the two matrix integral Eq.~(\ref{m-01}) by making use of the matrix integral representation
\begin{eqnarray}
\label{d-f-cv}
    \delta \left(
        {\boldsymbol A}
    \right) = \frac{1}{(2\pi)^{2n^2}} \int_{{\mathbb C}^{n\times n}} D{\boldsymbol q}\,
    \exp\left[\frac{i}{2} {\rm tr} \left(
            {\boldsymbol q} {\boldsymbol A}^\dagger + {\boldsymbol q}^\dagger {\boldsymbol A}
        \right)
    \right]
\end{eqnarray}
of the $\delta$-function of a general $n\times n$ complex valued matrix ${\boldsymbol A}$, where the flat measure $D{\boldsymbol q}$ is
\begin{eqnarray}\label{Dq}
    D{\boldsymbol q} = \prod_{\alpha=1}^n\prod_{\alpha^\prime=1}^n d\mathfrak{Re\,} q_{\alpha\alpha^\prime}\, d\mathfrak{Im\,} q_{\alpha\alpha^\prime}.
\end{eqnarray}
Setting
\begin{eqnarray}
{\boldsymbol A} = \tilde{{\boldsymbol {\mathcal W}}}-{\boldsymbol {\mathcal X}}{\boldsymbol {\mathcal Y}}^\dagger,
\end{eqnarray}
we rewrite Eq.~(\ref{m-01}) in the form
\begin{eqnarray}\fl
    d\pi_{n,m}(\tilde{{\boldsymbol {\mathcal W}}}) = \frac{1}{(2\pi)^{2n^2}}
    \int_{{\mathbb C}^{n\times n}} D{\boldsymbol q}\,
    \exp\left[ \frac{i}{2}\,{\rm tr} \left(
            {\boldsymbol q} {\tilde{\boldsymbol {\mathcal W}}}^\dagger + {\boldsymbol q}^\dagger {\tilde {\boldsymbol {\mathcal W}}}
        \right)
    \right] \nonumber\\
    \fl \qquad\qquad\times
    \int dP_{n,m}^{(2)}({\boldsymbol {\mathcal X}})\,
    \int dP_{n,m}^{(2)}({\boldsymbol {\mathcal Y}})
    \exp\left[
        - \frac{i}{2} {\rm tr} \left(
            {\boldsymbol q} {\boldsymbol {\mathcal Y}}{{\boldsymbol {\mathcal X}}}^\dagger +
            {\boldsymbol q}^\dagger {\boldsymbol {\mathcal X}} {{\boldsymbol {\mathcal Y}}}^\dagger
        \right)
    \right] \, D\tilde{{\boldsymbol {\mathcal W}}}.
\end{eqnarray}
By virtue of the identity
\begin{eqnarray}\label{g-id-01}
    \int dP_{n,m}^{(2)}({\boldsymbol {\mathcal Y}})\, \exp\left[
        i \,{\rm tr} \left(
            {\boldsymbol B} {\boldsymbol {\mathcal Y}} + {\boldsymbol C} {\boldsymbol {\mathcal Y}}^\dagger
        \right)
    \right] = \exp\left[
        -\frac{1}{a_2^\prime} {\rm tr} ({\boldsymbol B}{\boldsymbol C})
    \right]
\end{eqnarray}
with the matrices ${\boldsymbol B}$ and ${\boldsymbol C}$ of the size $m\times n$ and $n\times m$, respectively, being set to
\begin{eqnarray}
    {\boldsymbol B}={\boldsymbol C}^\dagger = -\frac{1}{2}{\boldsymbol {\mathcal X}}^\dagger {\boldsymbol q},
\end{eqnarray}
the integral over $dP_{n,m}^{(2)}({\boldsymbol {\mathcal Y}})$ can be performed to bring a reduced representation
\begin{eqnarray}\label{rep-red}\fl
    d\pi_{n,m}(\tilde{{\boldsymbol {\mathcal W}}}) = \frac{1}{(2\pi)^{2n^2}}
    \int_{{\mathbb C}^{n\times n}} D{\boldsymbol q}\,
    \exp\left[
         \frac{i}{2}\, {\rm tr} \left(
            {\boldsymbol q} {\tilde{\boldsymbol {\mathcal W}}}^\dagger + {\boldsymbol q}^\dagger {\tilde {\boldsymbol {\mathcal W}}}
        \right)
    \right] \nonumber\\
    \fl \qquad \qquad \qquad\times
    \int dP_{n,m}^{(2)}({\boldsymbol {\mathcal X}})\,
        \exp\left[
        -\frac{1}{4 a_2^\prime} {\rm tr} \left({\boldsymbol {\mathcal X}}^\dagger ({\boldsymbol q} {\boldsymbol q}^\dagger)
        {\boldsymbol {\mathcal X}}
        \right)
    \right]
     \, D\tilde{{\boldsymbol {\mathcal W}}}.
\end{eqnarray}
Applying the (Gaussian integral) identity
\begin{eqnarray} \label{g-id-02}
    \int dP_{n,m}^{(2)}({\boldsymbol {\mathcal X}})
    \, \exp\left[
        - {\rm tr} \left({\boldsymbol {\mathcal X}}^\dagger {\boldsymbol D}
        {\boldsymbol {\mathcal X}}
        \right)
    \right] = {\rm det}^{-m} \left(
    {\mathds 1}_n + \frac{1}{a_2} {\boldsymbol D}
    \right),
\end{eqnarray}
where the $n\times n$ matrix ${\boldsymbol D}$ is set to
\begin{eqnarray}
    {\boldsymbol D} = \frac{1}{4 a_2^\prime} \, {\boldsymbol q} {\boldsymbol q}^\dagger,
\end{eqnarray}
we calculate the integral over $dP_{n,m}^{(2)}({\boldsymbol {\mathcal X}})$ to arrive at the main result of this Subsection:
\begin{eqnarray}
\label{211-main}
    d\pi_{n,m}(\tilde{{\boldsymbol {\mathcal W}}}) &=& \frac{1}{(2\pi)^{2n^2}}
    \int_{{\mathbb C}^{n\times n}} D{\boldsymbol q}\,
    \exp\left[
         \frac{i}{2}\, {\rm tr} \left(
            {\boldsymbol q} {\tilde{\boldsymbol {\mathcal W}}}^\dagger + {\boldsymbol q}^\dagger {\tilde {\boldsymbol {\mathcal W}}}
        \right)
    \right] \nonumber\\
    &\times&
    {\rm det}^{-m} \left(
    {\mathds 1}_n + \frac{1}{4a_2 a_2^\prime} {\boldsymbol q}{\boldsymbol q}^\dagger
    \right)\, D\tilde{{\boldsymbol {\mathcal W}}}.
\end{eqnarray}
This matrix integral will be referred to as the type I representation of the probability measure $d\pi_{n,m}(\tilde{{\boldsymbol {\mathcal W}}})$ associated with non-Hermitean Wishart random matrices.

\subsubsection{Integral over positive definite Hermitean matrix (Type II)}\noindent\newline\newline
To derive an alternative representation of the probability measure $d\pi_{n,m}(\tilde{{\boldsymbol {\mathcal W}}})$, we start with
Eq.~(\ref{rep-red}) rewritten in the form
\begin{eqnarray}\label{rep-red-exchange}\fl
    d\pi_{n,m}(\tilde{{\boldsymbol {\mathcal W}}}) = \frac{1}{(2\pi)^{2n^2}}
    \int dP_{n,m}^{(2)}({\boldsymbol {\mathcal X}})\,
    \int_{{\mathbb C}^{n\times n}} D{\boldsymbol q}\,
    \exp\left[
         \frac{i}{2}\, {\rm tr} \left(
            {\boldsymbol q} {\tilde{\boldsymbol {\mathcal W}}}^\dagger + {\boldsymbol q}^\dagger {\tilde {\boldsymbol {\mathcal W}}}
        \right)
    \right] \nonumber\\
    \fl \qquad \qquad \qquad\times
        \exp\left[
        -\frac{1}{4 a_2^\prime} {\rm tr} \left({\boldsymbol q}^\dagger ({\boldsymbol {\mathcal X}}{\boldsymbol {\mathcal X}}^\dagger) {\boldsymbol q}
        \right)
    \right]
     \, D\tilde{{\boldsymbol {\mathcal W}}}.
\end{eqnarray}
Now we wish to integrate out the matrix ${\boldsymbol q}$. In the Wishart domain $m\ge n$, this can be achieved with the help of yet another easy-to-prove Gaussian integral identity
\begin{eqnarray}
    \int_{{\mathbb C}^{n\times n}} D{\boldsymbol q}\,
    \exp\left[
         i \, {\rm tr} \left(
            {\boldsymbol q} {\boldsymbol E}^\dagger + {\boldsymbol q}^\dagger {\boldsymbol E}
        \right)
    \right] \,        \exp\left[
        -{\rm tr} \left({\boldsymbol q} {\boldsymbol F} {\boldsymbol q}^\dagger)
        \right)
    \right]\nonumber \\
     \qquad\qquad=\pi^{n^2} {\rm det}^{-n} {\boldsymbol F}
        \exp\left[
            - {\rm tr\,}\left(
                {\boldsymbol E}^\dagger {\boldsymbol F}^{-1}
                {\boldsymbol E}\right)
        \right]
\end{eqnarray}
that holds for arbitrary $n\times n$ matrix ${\boldsymbol E}$ and a non-singular, positive-definite $n\times n$ matrix ${\boldsymbol F}$. Setting
\begin{eqnarray}
    {\boldsymbol E} = \frac{1}{2} \,{\tilde {\boldsymbol {\mathcal W}}},\quad {\boldsymbol F} = \frac{1}{4a_2^\prime} {\boldsymbol {\mathcal X}}{\boldsymbol {\mathcal X}}^\dagger,
\end{eqnarray}
we reduce Eq.~(\ref{rep-red-exchange}) to
\begin{eqnarray}\label{rr-2}
    d\pi_{n,m}(\tilde{{\boldsymbol {\mathcal W}}}) &=& \left(\frac{a_2^\prime}{\pi}\right)^{n^2}
    \int dP_{n,m}^{(2)}({\boldsymbol {\mathcal X}})\, {\rm det}^{-n}({\boldsymbol {\mathcal X}}{\boldsymbol {\mathcal X}}^\dagger)\nonumber \\
    &\times&
        \exp\left[
        -a_2^\prime\, {\rm tr} \left( \tilde{{\boldsymbol {\mathcal W}}}^\dagger ({\boldsymbol {\mathcal X}}{\boldsymbol {\mathcal X}}^\dagger)^{-1} \tilde{{\boldsymbol {\mathcal W}}}
        \right)
    \right]
     \, D\tilde{{\boldsymbol {\mathcal W}}}.
\end{eqnarray}
To facilitate integration over $dP_{n,m}^{(2)}({\boldsymbol {\mathcal X}})$, we insert an integrated $\delta$-function \footnote{Here, the flat measure $D{\boldsymbol Q}$ is
$$
    D{\boldsymbol Q} = \prod_{\alpha=1}^n dQ_{\alpha\alpha}\prod_{\alpha>\alpha^\prime=1}^n d\mathfrak{Re\,}Q_{\alpha,\alpha^\prime} \, d\mathfrak{Im\,}Q_{\alpha,\alpha^\prime}.
$$}
\begin{eqnarray}
    \int_{{\boldsymbol Q}^\dagger = {\boldsymbol Q}} D{\boldsymbol Q} \,\delta({\boldsymbol Q} - {\boldsymbol {\mathcal X}}{\boldsymbol {\mathcal X}}^\dagger)=1
\end{eqnarray}
into Eq.~(\ref{rr-2}) to come down to
\begin{eqnarray}\fl
    d\pi_{n,m}(\tilde{{\boldsymbol {\mathcal W}}}) =\left(\frac{a_2}{\pi}\right)^{nm} \left(\frac{a_2^\prime}{\pi}\right)^{n^2}
    \int_{{\boldsymbol Q}^\dagger ={\boldsymbol Q}} D{\boldsymbol Q}\, {\rm det}^{-n} {\boldsymbol Q}\,
     \exp\left[
        -a_2 \,{\rm tr\,}{\boldsymbol Q}
    \right]\nonumber\\
    \times \exp\left[-a_2^\prime\, {\rm tr\,}\left(  \tilde{{\boldsymbol {\mathcal W}}}^\dagger {\boldsymbol Q}^{-1} \tilde{{\boldsymbol {\mathcal W}}}
        \right)
    \right]\,
     \int_{{\mathbb C}^{n\times m}} D{\boldsymbol {\mathcal X}}\, \delta({\boldsymbol Q} - {\boldsymbol {\mathcal X}}{\boldsymbol {\mathcal X}}^\dagger)
     \, D\tilde{{\boldsymbol {\mathcal W}}}.
\end{eqnarray}
Finally, making use of the formula
\begin{eqnarray}\fl
     \int_{{\mathbb C}^{n\times m}} D{\boldsymbol {\mathcal X}}\, \delta({\boldsymbol Q} - {\boldsymbol {\mathcal X}}{\boldsymbol {\mathcal X}}^\dagger)
     = \pi^{n(2m-n+1)/2} \frac{\prod_{j=1}^{m-n} \Gamma(j)}{\prod_{j=1}^{m} \Gamma(j)}\, \Theta({\boldsymbol Q}) \, ({\rm det} {\boldsymbol Q} )^{m-n}
\end{eqnarray}
proven in Appendix \ref{app-2}, we end up with the sought alternative (type II) representation of the probability measure $d\pi_{n,m}(\tilde{{\boldsymbol {\mathcal W}}})$ for non-Hermitean Wishart random matrices:
\begin{eqnarray}\fl \label{212-main-a}
    d\pi_{n,m}(\tilde{{\boldsymbol {\mathcal W}}}) =
    \frac{(a_2 a_2^\prime)^{n(n+m)/2}}{\pi^{n(3n-1)/2}} \frac{\prod_{j=1}^{m-n} \Gamma(j)}{\prod_{j=1}^{m} \Gamma(j)}
    \nonumber\\
    \fl \qquad\times
    \int_{{\boldsymbol Q}^\dagger ={\boldsymbol Q}} D{\boldsymbol Q}\, \Theta({\boldsymbol Q})\, {\rm det}^{m-2n} {\boldsymbol Q}\,
     \exp\left[-\sqrt{a_2 a_2^\prime} \,{\rm tr\,}( {\boldsymbol Q} + \tilde{{\boldsymbol {\mathcal W}}}^\dagger {\boldsymbol Q}^{-1} \tilde{{\boldsymbol {\mathcal W}}})
    \right]\, D\tilde{{\boldsymbol {\mathcal W}}}.\nonumber\\
    {}
\end{eqnarray}
Equation (\ref{212-main-a}) represents the main result of this Subsection.
\subsection{Spectral statistics}
To study the spectral statistics of non-Hermitean Wishart random matrices, we need to evaluate the joint probability density function (j.p.d.f.)
$\rho_{n,m}(w_1,\dots,w_n)$ of all $n$ complex eigenvalues of $\tilde{\boldsymbol {\mathcal W}}$. Although both type I and type II matrix integral representations [Eqs.~(\ref{211-main}) and (\ref{212-main-a})] of the probability measure $d\pi_{n,m}(\tilde{{\boldsymbol {\mathcal W}}})$ can be used to this end, the derivation based on
Eqs.~(\ref{211-main}) is particularly elegant.

\subsubsection{Joint probability density function of all eigenvalues}\label{Sec-2-2-1}
\noindent\newline\newline
On general grounds, the probability measure $d\rho_{n,m}(w_1,\dots,w_n)$ in the complex eigenvalue space is related to the probability measure $d\pi_{n,m}(\tilde{{\boldsymbol {\mathcal W}}})$ via the integral transformation
\begin{eqnarray}\fl\label{presc}
    d\rho_{n,m}(w_1,\dots,w_n) = \int_{{\boldsymbol U}\in{\mathbb U}(n)}  \int_{{\boldsymbol R} \in {\mathbb C}^{n(n-1)/2}}
    d\pi_{n,m}\left(\tilde{{\boldsymbol {\mathcal W}}} = {\boldsymbol U} ({\boldsymbol w} + {\boldsymbol R}) {\boldsymbol U}^\dagger\right).
\end{eqnarray}
This prescription is a consequence of the Schur decomposition of the $n\times n$ complex matrix $\tilde{{\boldsymbol {\mathcal W}}}$ with distinct eigenvalues \footnote{~See Appendix 33 in Mehta (2004).}
\begin{eqnarray}
\label{schur}
        \tilde{{\boldsymbol {\mathcal W}}} = {\boldsymbol U}\, ({\boldsymbol w} + {\boldsymbol R})\,{\boldsymbol U}^\dagger,
\end{eqnarray}
where ${\boldsymbol U}$ is a unitary matrix, ${\boldsymbol R}$ is strictly upper triangular,
\begin{eqnarray}
    ({\boldsymbol R})_{jk} = \left\{
                               \begin{array}{cc}
                                 R_{jk}\in {\mathbb C}, & \hbox{$j<k$}, \\
                                 0, & \hbox{$j\ge k$},
                               \end{array}
                             \right.
\end{eqnarray}
and ${\boldsymbol w}={\rm diag}(w_1,\dots,w_n)$ is a diagonal matrix composed of $n$ complex eigenvalues of $\tilde{{\boldsymbol {\mathcal W}}}$. This decomposition is unique if we label the eigenvalues and require the first nonzero element in each column of ${\boldsymbol U}$ to be positive. It holds that
\begin{eqnarray}
    D{\tilde{\boldsymbol {\mathcal W}}} = D{\boldsymbol W}
     \,  D{\boldsymbol R}\, d\mu({\boldsymbol U}),
\end{eqnarray}
where
\begin{eqnarray} \label{dw}
    D{\boldsymbol W} &=& |\Delta_n({\boldsymbol w})|^2 \prod_{j=1}^n d\mathfrak{Re}\, w_j \, d\mathfrak{Im}\, w_j,\\
    D{\boldsymbol R} &=& \prod_{j<k=1}^n d\mathfrak{Re}\, {R}_{jk}
    \, d\mathfrak{Im}\, {R}_{jk},
\end{eqnarray}
and $d\mu({\boldsymbol U})$ denotes the Haar measure on ${\mathbb U}(n)$.
\noindent\newline\newline
Below, the integrals in Eq.~(\ref{presc}) will be evaluated based on the type I representation
\begin{eqnarray}
\label{211-restart}
    d\pi_{n,m}(\tilde{{\boldsymbol {\mathcal W}}}) &=& \frac{1}{(2\pi)^{2n^2}}
    \int_{{\mathbb C}^{n\times n}} D{\boldsymbol q}\,
    \exp\left[
         \frac{i}{2}\, {\rm tr} \left(
            {\boldsymbol q} {\tilde{\boldsymbol {\mathcal W}}}^\dagger + {\boldsymbol q}^\dagger {\tilde {\boldsymbol {\mathcal W}}}
        \right)
    \right] \nonumber\\
    &\times&
    {\rm det}^{-m} \left(
    {\mathds 1}_n + \frac{1}{4a_2 a_2^\prime} {\boldsymbol q}{\boldsymbol q}^\dagger
    \right)\, D\tilde{{\boldsymbol {\mathcal W}}}
\end{eqnarray}
of the probability measure $d\pi_{n,m}(\tilde{{\boldsymbol {\mathcal W}}})$. We shall proceed in three steps.
\newline\newline\noindent
Step $\natural\; 1$.---Substitute the Schur decomposed matrix $\tilde{{\boldsymbol {\mathcal W}}}$ [Eq.~(\ref{schur})] into Eq.~(\ref{211-restart}) to observe that
\begin{eqnarray}\fl
        d\rho_{n,m}(w_1,\dots,w_n) = \frac{{\rm vol}[{\mathbb U}(n)]}{(2\pi)^{2n^2}} \, D{\boldsymbol W}\,
        \nonumber\\
        \times
        \int_{{\mathbb C}^{n\times n}} D{\boldsymbol \xi}\,
    \exp\left[
         \frac{i}{2}\, {\rm tr} \left(
            {\boldsymbol \xi} {\boldsymbol w}^\dagger + {\boldsymbol \xi}^\dagger {\boldsymbol w}
        \right)
    \right] {\rm det}^{-m} \left(
    {\mathds 1}_n + \frac{1}{4a_2 a_2^\prime} {\boldsymbol \xi}{\boldsymbol \xi}^\dagger
    \right)\nonumber\\
    \times \int_{{\boldsymbol R} \in{\mathbb C}^{n(n-1)/2}} D{\boldsymbol R}\,
    \exp\left[
         \frac{i}{2}\, {\rm tr} \left(
            {\boldsymbol \xi} {\boldsymbol R}^\dagger + {\boldsymbol \xi}^\dagger {\boldsymbol R}
        \right)
    \right].
\end{eqnarray}
Here, ${\boldsymbol \xi} = {\boldsymbol U}^\dagger {\boldsymbol q} \, {\boldsymbol U}$ is a new integration matrix, and
\begin{eqnarray}
    {\rm vol}[{\mathbb U}(n)] = \int d\mu({\boldsymbol U}) = \frac{\pi^{n(n-1)/2}}{n! \prod_{j=1}^n \Gamma(j)}
\end{eqnarray}
is the volume of the unitary group ${\mathbb U}(n)$. Notice that the integral over `radial' degrees of freedom $D{\boldsymbol R}$ has factored out.
\newline\newline\noindent
Step $\natural\; 2$.---Perform the radial integral
\begin{eqnarray}\fl
    \int_{{\boldsymbol R} \in{\mathbb C}^{n(n-1)/2}} D{\boldsymbol R}\,
    \exp\left[
         \frac{i}{2}\, {\rm tr} \left(
            {\boldsymbol \xi} {\boldsymbol R}^\dagger + {\boldsymbol \xi}^\dagger {\boldsymbol R}
        \right)
    \right] = (2\pi)^{n(n-1)}
      \prod_{j<k=1}^n\, \delta^{(2)} (\xi_{jk}),
\end{eqnarray}
to find out the remarkable formula
\begin{eqnarray}\fl
        d\rho_{n,m}(w_1,\dots,w_n) = \frac{{\rm vol}[{\mathbb U}(n)]}{(2\pi)^{n(n+1)}} \, D{\boldsymbol W}\,
        \nonumber\\
        \fl \qquad \times
        \int_{{\boldsymbol \psi} \in {\mathbb C}^{n(n+1)/2}} D{\boldsymbol \psi}\,
    \exp\left[
         \frac{i}{2}\, {\rm tr} \left(
            {\boldsymbol \psi} {\boldsymbol w}^\dagger + {\boldsymbol \psi}^\dagger {\boldsymbol w}
        \right)
    \right] {\rm det}^{-m} \left(
    {\mathds 1}_n + \frac{1}{4a_2 a_2^\prime} {\boldsymbol \psi}{\boldsymbol \psi}^\dagger
    \right),
\end{eqnarray}
where ${\boldsymbol \psi}$ is an $n\times n$ complex valued lower triangular matrix,
\begin{eqnarray}
    ({\boldsymbol \psi})_{jk} = \left\{
                               \begin{array}{cc}
                                 0, & \hbox{$j<k$}, \\
                                 \psi_{jk}\in {\mathbb C}, & \hbox{$j\ge k$.}
                               \end{array}
                             \right.
\end{eqnarray}
\newline\noindent
Step $\natural\; 3$.---Use Lemma 3 of Appendix \ref{app-2} to eventually derive:
\begin{eqnarray} \fl
    d\rho_{n,m}(w_1,\dots,w_n) = \frac{2^n (a_2 a_2^\prime)^{n(m+1)/2}}{\pi^n n!} \frac{\prod_{j=1}^{m-n}\Gamma(j)}{\prod_{j=1}^n \Gamma(j) \prod_{j=1}^m \Gamma(j)}
    \nonumber\\
    \fl \qquad \qquad \qquad\qquad \times
    \prod_{j=1}^n |w_j|^{m-n} \, K_{m-n}\left(
    2 |w_j| \sqrt{a_2 a_2^\prime}
    \right)\, D{\boldsymbol W}.
\end{eqnarray}
Considered together with Eq.~(\ref{dw}), this completes our calculation of the j.p.d.f.
\begin{eqnarray} \fl \label{rho-nh}
    \rho_{n,m}(w_1,\dots,w_n) = \frac{2^n (a_2 a_2^\prime)^{n(m+1)/2}}{\pi^n n!} \frac{\prod_{j=1}^{m-n}\Gamma(j)}{\prod_{j=1}^n \Gamma(j) \prod_{j=1}^m \Gamma(j)}
    \nonumber\\
    \fl \qquad \qquad \qquad\qquad \times
    |\Delta_n({\boldsymbol w})|^2 \, \prod_{j=1}^n |w_j|^{m-n} \, K_{m-n}\left(
    2 |w_j| \sqrt{a_2 a_2^\prime}
    \right)
\end{eqnarray}
of all $n$ complex eigenvalues for a non-Hermitean Wishart random matrix model defined
by Eqs.~(\ref{eq-12nh}) -- (\ref{eq-12a-nhw-2}) in Section \ref{Sec-1-2-1}. Equation~(\ref{rho-nh}) represents the main result of this Subsection. It can be viewed as a non-Hermitean counterpart of Eq.~(\ref{wish-jpdf}) considered at $\beta=2$. We note that a closely related result was first obtained by Osborn (2004) in the context of QCD physics, see also Akemann, Osborn, Splittorff and Verbaarschot (2005).

\subsubsection{Correlation functions}\label{Sec-2-2-2}
\noindent\newline\newline
The $p$-point correlation function is defined in the standard manner (Mehta 2004)
\begin{eqnarray}
\label{rk=2} \fl
    R_{n,m}^{(p)}(w_1,\cdots,w_p) = \frac{n!}{(n-p)!} \prod_{j=p+1}^n\int_{\mathbb C} d\mathfrak{Re}\, w_j \, d\mathfrak{Im}\, w_j\, \rho_{n,m}(w_1,\cdots,w_p,w_{p+1},\cdots,w_n).
    \nonumber\\
{}
\end{eqnarray}
Here,
\begin{eqnarray}\label{rho-jpdf}
    \rho_{n,m}(w_1,\cdots,w_n) = c_{n,m}^{-1} \, |\Delta_n({\boldsymbol w})|^2 \prod_{j=1}^n |w_j|^{\nu} K_{\nu}(|w_j|)
\end{eqnarray}
and
\begin{eqnarray}
        c_{n,m} = 2^{nm} \pi^n n!\, \frac{\prod_{j=1}^n \Gamma(j) \prod_{j=1}^m \Gamma(j)}{\prod_{j=1}^{m-n}\Gamma(j)}.
\end{eqnarray}
The parameter $\nu$ is defined in Eq.~(\ref{m-nu}).  [In order to simplify notation, we have set $a_2 a_2^\prime = 1/4$ in Eq.~(\ref{rho-nh})].
\noindent\newline\newline
To integrate out $n-p$ eigenvalues in Eq.~(\ref{rk=2}), we employ the orthogonal polynomials technique (Mehta 2004). Introducing a set of monic polynomials
\begin{eqnarray}
    p_j(w) = w^j + a_1 w^{j-1} + \cdots
\end{eqnarray}
orthogonal on ${\mathbb C}$ with respect to the measure
\begin{eqnarray}
    d\mu(w) = |w|^{\nu} K_{\nu}(|w|)\, d\mathfrak{Re}\, w \,\, d\mathfrak{Im}\, w,
\end{eqnarray}
that is,
\begin{eqnarray}
    \int_{\mathbb C} d\mu(w)\, p_j(w) \, p_k(\bar{w}) ={\mathcal N}_j \delta_{jk}.
\end{eqnarray}
Rotational symmetry of the weight function in the measure $d\mu(w)$ suggests that $p_j(w)$'s are merely the monomials
\begin{eqnarray}
    p_j(w) = w^j
\end{eqnarray}
with the normalisation constant ${\mathcal N}_j$ being \footnote[1]{Notice that $$\prod_{j=0}^{n-1} {\mathcal N}_j = \frac{c_{n,m}}{n!}$$ which implies, through the Andr\'eief-de Bruijn integration formula (Andr\'eief 1883 and de Bruijn 1955) that the probability measure Eq.~(\ref{rho-jpdf}) is properly normalised:
\begin{eqnarray}
    \fl \qquad
    \prod_{j=1}^n\int_{\mathbb C} d\mathfrak{Re}\, w_j \, d\mathfrak{Im}\, w_j\, \rho_{n,m}(w_1,\cdots,w_n)\nonumber\\
    \fl \qquad\qquad\qquad = \frac{n!}{c_{n,m}}
    \, {\rm det} \left(
        \int_{\mathbb C} d\mathfrak{Re}\, w \, d\mathfrak{Im}\, w\, w^j \bar{w}^k \, |w|^{\nu} K_{\nu}(|w|)
    \right)
    = \frac{n!}{c_{n,m}} \, \prod_{j=0}^{n-1} {\mathcal N}_j = 1. \nonumber
\end{eqnarray}}
\begin{eqnarray}
    {\mathcal N}_j = \int_{\mathbb C} d\mu(w)\, |w|^{2j} = 2^{2j+\nu+1} \pi \Gamma(j+1) \Gamma(j+\nu+1).
\end{eqnarray}
This observation allows us to rewrite the j.p.d.f. Eq.~(\ref{rho-jpdf}) in the form
\begin{eqnarray} \fl
    \rho_{n,m}(w_1,\cdots,w_n) = \frac{1}{n!} \, {\rm det}\left[\frac{p_{k-1}(w_j)}{\sqrt{{\mathcal N}_{k-1}}}\right]
    {\rm det}\left[\frac{p_{k-1}(\bar{w}_j)}{\sqrt{{\mathcal N}_{k-1}}}\right]
    \,\prod_{j=1}^n |w_j|^{\nu} K_{\nu}(|w_j|).\nonumber \\
    {}
\end{eqnarray}
Equivalently,
\begin{eqnarray} \label{rho-det}
    \rho_{n,m}(w_1,\cdots,w_n) = \frac{1}{n!} \, {\rm det}\left[
            {\mathbb K}_{n,m}(w_j, \bar{w}_k)
\right],
\end{eqnarray}
where
\begin{eqnarray}\fl \label{ex-kernel}
    {\mathbb K}_{n,m}(w,w^\prime)= \frac{1}{2^{\nu+1} \pi} \left|w w^\prime \right|^{\nu/2} \,\sqrt{K_{\nu}(|w|) K_{\nu}(|w^\prime|)}
    \sum_{k=0}^{n-1} \frac{(w w^\prime)^k} {2^{2k} k!\, (k+\nu)!}\nonumber\\
    {}
\end{eqnarray}
is the scalar kernel that obeys the projection property
\begin{eqnarray}
\label{proj}
    \int_{\mathbb C} d\mathfrak{Re}\, \xi \, d\mathfrak{Im}\, \xi\, \,{\mathbb K}_{n,m}(w,\bar{\xi})\, {\mathbb K}_{n,m}(\xi,w^\prime) = {\mathbb K}_{n,m}(w,w^\prime).
\end{eqnarray}
By virtue of the Dyson integration theorem (Mehta 1976), Eqs.~(\ref{rk=2}), (\ref{rho-det}) and (\ref{proj}) imply that the $p$-point correlation function equals
\begin{eqnarray}
\label{rp}
    R_{n,m}^{(p)}(w_1,\cdots,w_p) =
    \, {\rm det} \left[
        {\mathbb K}_{n,m}(w_j, \bar{w}_k)
    \right]_{1\le j,k \le p}
\end{eqnarray}
as was anticipated. Note that a closely related result appears in the recent work by Akemann, Phillips and Shifrin (2009).

\subsection{Large-$n$/large-$\nu$ analysis of the mean density of eigenvalues}\label{Sec-2-3}
Equations (\ref{ex-kernel}) and (\ref{rp}) show that the mean density of complex eigenvalues is determined by the formula
\begin{eqnarray}\label{ex-density}
    R_{n,m}^{(1)}(w) = {\mathbb K}_{n,m}(w, \bar{w}) = \frac{1}{2^{\nu+1}\pi} \,|w|^\nu\, K_{\nu}(|w|) \, {\mathcal F}_{n,\nu}(|w|),
\end{eqnarray}
where the function ${\mathcal F}_{n,\nu}(\varrho)$ equals
\begin{eqnarray}
\label{f-ex-density}
{\mathcal F}_{n,\nu}(\varrho) =
\sum_{k=0}^{n-1} \frac{1}{k!\, (k+\nu)!}
    \left( \frac{\varrho}{2}\right)^{2k},\qquad \varrho \in {\mathbb R}.
\end{eqnarray}
Equation~(\ref{ex-density}) is exact as it is valid for arbitrary $m\ge n$.  Below, we shall be interested in its large-$n$/large-$\nu$ analysis.
\newline\noindent
\subsubsection{The case $n$ fixed and $\nu\gg 1$ (Regime I)}
\noindent\newline\newline
This region of parameters corresponds to meticulously sampled time series. To study the asymptotic behaviour of the mean density of eigenlevels [Eq.~(\ref{ex-density})] in this regime, we focus on the function ${\mathcal F}_{n,\nu}(\varrho)$ defined by Eq.~(\ref{f-ex-density}) to realise that, for  $\nu\gg 1$, the inverse factorial $[(k+\nu)!]^{-1}$ therein can be approximated by the leading term of the Stirling formula,
\begin{eqnarray}
    \frac{1}{(k+\nu)!} \simeq \frac{1}{\sqrt{2\pi}}\,\frac{e^\nu}{\nu^{\nu+1/2}} \frac{1}{\nu^k},\qquad \nu \gg 1.
\end{eqnarray}
Consequently, we derive
\begin{eqnarray}\fl
    {\mathcal F}_{n,\nu}(\varrho)\Big|_{\nu \gg 1} \simeq \frac{1}{\sqrt{2\pi}} \frac{e^\nu}{\nu^{\nu+1/2}} \sum_{k=0}^{n-1} \frac{1}{k!}
    \left( \frac{\varrho^2}{4\nu}\right)^{k}
    =
    \frac{1}{\sqrt{2\pi}}\frac{e^\nu}{\nu^{\nu+1/2} \Gamma(n)}\,  e^{\varrho^2/4\nu} \,\Gamma\left(
    n, \displaystyle{\frac{\varrho^2}{4\nu}}
    \right).\nonumber\\
    {}
\end{eqnarray}
This result prompts that for the large-$\nu$ limit of the spectral density to be well defined, the energy variable $|w|$ must scale with $\sqrt{\nu}$ so that their ratio
\begin{eqnarray}
    x = \frac{|w|}{\sqrt{\nu}} \sim \mathcal{O}(\nu^0)
\end{eqnarray}
is kept bounded. Having appreciated this fact, we make use of Lemma 4 of the Appendix \ref{app-3} to treat the modified Bessel function $K_\nu(|w|)$ in Eq.~(\ref{ex-density}). As the result, Eq.~(\ref{ex-density}) boils down to the elegant expression
\begin{eqnarray}\label{reg-2-n}
    R_{n,m}^{(1)}(w) = \frac{1}{4\pi \nu\, \Gamma(n)} \,\Gamma\left(
    n, \displaystyle{\frac{|w|^2}{4\nu}}
    \right).
\end{eqnarray}
Equation~(\ref{reg-2-n}) holds for $n$ finite and $\nu \gg 1$. A detailed discussion of the qualitative behaviour of the mean eigenvalue density Eq.~(\ref{reg-2-n}) has been presented in Section \ref{Sec-1-2-2}, see also Fig. \ref{dos-regime-01} (right panel).
\newline\noindent
\subsubsection{The case $n\gg 1$ and $\nu\ge 0$ fixed (Regime II)}\label{Sec-2-3-1}
\noindent\newline\newline
For $\nu \sim {\mathcal O}(n^0)$, two different large-$n$ subcases should be distinguished.\noindent\newline
\begin{itemize}
  \item If $|w|\sim {\mathcal O}(n^0)$, the upper summation limit in Eq.~(\ref{f-ex-density}) can be extended to infinity yielding \footnote{We have used the Taylor series
$$
    I_\nu(z) = \left( \frac{z}{2}\right)^\nu \sum_{k=0}^{\infty} \frac{1}{k!\, (k+\nu)!}
    \left( \frac{z}{2}\right)^{2k}.
$$
[See \texttt{http://functions.wolfram.com/03.02.02.0001.01}].}
    \begin{eqnarray}\label{r-bulk}
        R_{\infty}^{(1)}(w) = \frac{1}{2\pi} \, I_\nu(|w|)\, K_\nu(|w|),
    \end{eqnarray}
where $I_\nu(|w|)$ denotes the modified Bessel function of the first kind. In particular, this simple expression is useful in the analysis of the large-$n$ mean density $R_1(|w|)$ both close to the origin ($|w|\ll 1$) \footnote{This spectral region is of particular interest in QCD physics, see e.g. Akemann, Phillips and Shifrin (2009).} and far away from it ($|w| \gg 1$).
\noindent\newline\newline
Close to the origin ($|w|\ll 1$), we obtain \footnote{In addition to the Taylor series for $I_\nu(z)$, we have also used the small-$z$ expansion
$$
    K_\nu(z) = \left\{
               \begin{array}{ll}
                 \displaystyle{-\log\left(\frac{z}{2} \right)-\gamma}, & \hbox{$\nu = 0$,} \\
                    {} & {}\\
                 \displaystyle{\frac{\Gamma(\nu)}{2}\, \left( \frac{2}{z} \right)^\nu}, & \hbox{$\nu \in {\mathbb Z}_+$,}
               \end{array}
             \right.
$$
exhibiting the main terms as $z\rightarrow 0$. Here, $\gamma=0.577215665\dots$ is the Euler constant. [See \texttt{http://functions.wolfram.com/03.04.06.0044.01}].}:
\begin{eqnarray}
\label{r-zero}
    R_\infty^{(1)}(w)\Big|_{|w|\ll 1} = \left\{
               \begin{array}{ll}
                 \displaystyle{\frac{1}{2\pi} \log \frac{1}{|w|}}, & \hbox{$\nu=0$,}\\
                    {} & {}\\
                 \displaystyle{\frac{1}{4\pi \nu}}, & \hbox{$\nu \neq 0$.}
               \end{array}
             \right.
\end{eqnarray}
In the opposite limit, the unusual power-law decay of the mean density is observed \footnote{For $|{\rm arg} (z)|<\pi/2$ and $|z|\rightarrow \infty$, it holds:
$$
    I_\nu(z) \simeq \frac{1}{\sqrt{2\pi z}} \, e^z \left(
    1+{\mathcal O}\left( \frac{1}{z}
\right)
\right),
$$
and
$$
    K_\nu(z) \simeq \sqrt{\frac{\pi}{2 z}} \, e^{-z} \left(
    1+{\mathcal O}\left( \frac{1}{z}
\right)
\right).
$$
[See \texttt{http://functions.wolfram.com/03.02.06.0006.01} and \texttt{../03.04.06.0010.01}.]
}:
\begin{eqnarray}
\label{r-large}
    R_\infty^{(1)}(w)\Big|_{|w|\gg 1} = \frac{1}{4\pi |w|}.
\end{eqnarray}
Both laws [Eqs.~(\ref{r-zero}) and (\ref{r-large})] are in sharp contrast with the mean density of complex eigenvalues in the GinUE (Ginibre's unitary ensemble), a close relative of the non-Hermitean Wishart matrix model. In the former ensemble, both domains ($|w|\ll 1$ and $|w|\gg 1$) are characterised by the constant mean density (Ginibre 1965)
\begin{eqnarray}
    R^{(1,\rm GinUE)}_{\infty}(w) \simeq \frac{1}{\pi}
\end{eqnarray}
indicative of the emerging Girko law (Girko 1984, Girko 1986, Bai 1997).
\noindent\newline\newline
The slow decay at infinity exhibited by Eq.~(\ref{r-large}) implies that this limiting law must break down at some `critical' point $|w|=|w_c|$. The mean density normalisation
\begin{eqnarray}
    2\pi \int_0^{|w_c|} d|w| \, |w| \times \frac{1}{4\pi |w|} =n
\end{eqnarray}
suggests that
\begin{eqnarray}\label{crit-point}
|w_c| \simeq 2n.
\end{eqnarray}
The position of the break-down point Eq.~(\ref{crit-point}) is clearly seen in Fig. \ref{dos-regime-02} (right panel). \noindent\newline\newline
{\it Remark.}---We note in passing that the algebraically decaying large-$n$ density Eq.~(\ref{r-large}) as well as the position of the critical point $|w_c|=2n$ can alternatively be derived within the two-dimensional Coulomb fluid approach (Chau and Zaboronsky 1998, Zabrodin and Wiegmann 2006). Indeed, given the j.p.d.f. of the form
\begin{eqnarray}
\rho_n(w_1,\dots,w_n) \propto |\Delta_n({\boldsymbol w})|^2 \prod_{j=1}^n e^{-V(w_j,\bar{w}_j)},
\end{eqnarray}
it is straightforward to realise that, in the leading order in $1/n$, the mean eigenvalue density
\begin{eqnarray}\label{dos-2d-cf}
R_n^{(1)}(w) = \frac{\boldsymbol{\Delta}}{4\pi} \, V(w,\bar{w})
\end{eqnarray}
is proportional to the Laplacian
\begin{eqnarray}
    \boldsymbol{\Delta} = 4 \frac{\partial^2}{\partial w \partial \bar{w}}
\end{eqnarray}
of the confinement potential $V(w,\bar{w})$. For confinement potentials of the form $V(w,\bar{w}) = V(|w|)$, the eigenvalue density is supported within the disk of the radius $R=|w_c|$, where $|w_c|$ is a (positive) solution of the equation
\begin{eqnarray}\label{dos-R-cf}
    \frac{1}{2}\left(
        R \frac{\partial V(R)}{\partial R} - \lim_{r\rightarrow 0} r \frac{\partial V(r)}{\partial r}
    \right) = n.
\end{eqnarray}
It is a simple exercise to verify that the large-$|w|$ expansion of the modified Bessel function $K_\nu(|w|)$ in Eq.~(\ref{rho-jpdf}) yields the effective confinement potential of the form
\begin{eqnarray}
    V_{\rm eff}(|w|\gg 1) = |w| -\left(\nu-\frac{1}{2}\right) \,\log |w| + {\mathcal O}(|w|^{-1}).
\end{eqnarray}
Applying Eqs.~(\ref{dos-2d-cf}) and (\ref{dos-R-cf}), one reproduces Eqs. (\ref{r-large}) and (\ref{crit-point}), respectively.  $\;\;\;\blacksquare$
\noindent\newline
  \item To probe the mean eigenlevel density in the vicinity of the critical point $|w_c|\simeq 2n$, we have to analyse the exact formulae Eqs.~(\ref{ex-density}) and (\ref{f-ex-density}) for $|w| \sim {\mathcal O}(n^1)$. We found it useful to put forward the multiplicative ansatz
\begin{eqnarray}
    R_{n,m}^{(1)}(w) = R_\infty^{(1)}(w) \cdot {\mathbb T}_c(|w|),
\end{eqnarray}
where the function ${\mathbb T}_c(|w|)={\mathbb T}_c(|w|;n,\nu)$ is to account for (anticipated) significant deviations of the mean density of eigenlevels from the `na\"{i}ve' limiting curve $R_\infty^{(1)}(w)$ given by Eq.~(\ref{r-bulk}) in the vicinity of the critical point $|w_c|\simeq 2n$. In the region $1\ll |w| \ll |w_c|$, the function ${\mathbb T}_c(|w|)$ is
expected to approach unity.
\newline\newline\noindent
To determine ${\mathbb T}_c(\varrho)$ for $\varrho\in {\mathbb R}_+$, we first spot that yet another function ${\mathcal F}_{n,\nu}(\varrho)$ [Eq.~(\ref{f-ex-density})]
entering the exact solution Eq.~(\ref{ex-density}), satisfies the differential equation
\begin{eqnarray}\fl\label{feq}
{\mathcal F}_{n,\nu}^{\prime\prime}(\varrho) + \frac{2\nu+1}{\varrho} \, {\mathcal F}_{n,\nu}^{\prime}(\varrho) - {\mathcal F}_{n,\nu}(\varrho) = -  \frac{1}{\Gamma(n)\, \Gamma(n+\nu)}
    \left( \frac{\varrho}{2}\right)^{2(n-1)}.
\end{eqnarray}
Owing to the relation
\begin{eqnarray}
    {\mathcal F}_{n,\nu}(\varrho) = \left(
            \frac{2}{\varrho}
    \right)^\nu  I_\nu(\varrho) \,
    {\mathbb T}_c(\varrho),
\end{eqnarray}
the function ${\mathbb T}_c(\varrho)$ is seen to satisfy the differential equation
\begin{eqnarray} \fl \label{tc-exact}
    {\mathbb T}_c^{\,\prime\prime}(\varrho)  + \frac{1}{\varrho} \left[
        1+ 2 \varrho \frac{I_\nu^\prime(\varrho)}{I_\nu(\varrho)}
    \right]\,{\mathbb T}_c^{\,\prime}(\varrho) = -  \frac{1}{\Gamma(n)\, \Gamma(n+\nu)}
    \frac{1}{I_\nu(\varrho)}\left( \frac{\varrho}{2}\right)^{2(n-1)+\nu}.\nonumber\\
{}
\end{eqnarray}
By derivation, this equation is {\it exact} as it is valid for arbitrary $\nu\ge 0$, $n$ and $\varrho \in {\mathbb R}$. As soon as we are interested in the large-$n$ analysis of a vicinity of the critical point $|w| \sim |w_c| \simeq 2n$, the large-$\varrho$ reduction of Eq.~(\ref{tc-exact}) matters. For $\varrho\gg 1$, the above equation simplifies to
\begin{eqnarray} \label{tc-appr}
    {\mathbb T}_c^{\,\prime\prime}(\varrho)  + 2\,{\mathbb T}_c^{\,\prime}(\varrho) = - c_{n,\nu} \, \varrho^{2n+\nu-3/2} \, e^{-\varrho},
\end{eqnarray}
where
\begin{eqnarray}
    c_{n,\nu} = \frac{\sqrt{\pi}}{2^{2n+\nu-5/2} \Gamma(n) \Gamma(n+\nu)} \simeq \frac{e^{2n}}{\sqrt{\pi n}} \frac{1}{(2n)^{2n+\nu-3/2}}.
\end{eqnarray}
Focussing on an $n^\alpha$-vicinity of the critical point by setting
\begin{eqnarray}
    |w|=\varrho = 2n + 2 n^\alpha t,\qquad t\sim {\mathcal O}(n^0),
\end{eqnarray}
where the exponent $\alpha$ is yet to be determined, we introduce the new function
\begin{eqnarray}
    {\mathbb A}_c(t) = {\mathbb T}_c(\varrho=2n + 2 n^\alpha t;n,\nu)
\end{eqnarray}
which appears to have a well-defined large-$n$ limit provided $\alpha=1/2$. This limit is described by the equation
\begin{eqnarray}
    {\mathbb A}_c^\prime (t) = - \frac{1}{\sqrt{\pi}} \, e^{-t^2}.
\end{eqnarray}
Its solution, that approaches unity at $t \rightarrow -\infty$ (that is, deep in the bulk to the left of the critical point) reads:
\begin{eqnarray}
    {\mathbb A}_c(t) = \frac{1}{2}{\rm erfc\,}(t).
\end{eqnarray}
Equivalently,
\begin{eqnarray}
    {\mathbb T}_c(\varrho) = \frac{1}{2}{\rm erfc\,}\left(\frac{\varrho-2n}{2\sqrt{n}}\right).
\end{eqnarray}
Hence, we conclude that the mean density of eigenlevels in the $\sqrt{n}$-vicinity \footnote{Put differently, Eq.~(\ref{crit-vic}) stays valid for
$$
\frac{|w|-2n}{2\sqrt{n}} \sim {\mathcal O}(n^0)
$$
kept bounded.} of the critical point is given by the formula
\begin{eqnarray}
\label{crit-vic}
    R_{n,m}^{(1)}(w) = \frac{1}{8\pi |w|}\, {\rm erfc\,}\left(\frac{|w|-2n}{2\sqrt{n}}\right).
\end{eqnarray}
Notice that Eq.~(\ref{crit-vic}) predicts that the mean density of eigenlevels at the critical point is two times smaller as compared to the na\"ive result Eq.~(\ref{r-large}) if extended down to $|w|=|w_c|$. This nonperturbative piece of the mean density is responsible for forming a `goblet base' in Fig. \ref{dos-regime-02} (right panel). \newline
\end{itemize}
The overall behaviour of the mean density is well captured by the single formula
  \begin{eqnarray}\label{d-global-sec}
        R_{n,m}^{(1)}(w) \simeq \frac{1}{4\pi}\, I_\nu(|w|) \, K_\nu(|w|) \, {\rm erfc\,}\left(\frac{|w|-2n}{2\sqrt{n}}\right)
  \end{eqnarray}
as has been discussed in Section \ref{Sec-1-2-2}, see also Fig. \ref{dos-regime-02} (right panel).

\subsubsection{The case $n\gg 1$ and $\nu/n=q > 0$ fixed (Regime III)}\label{Sec-2-3-3}\noindent\newline\newline
This region of parameters corresponds to the Mar\u{c}enko--Pastur domain discussed in Section \ref{Sec1-1}. To study the large-$n$ behaviour of the mean density of complex eigenlevels, we follow the idea introduced in Section \ref{Sec-2-3-1}. Specifically, we decompose the mean density as
\begin{eqnarray}
\label{ans-2}
    R_{n,m}^{(1)}(w) = R_\infty^{(1)}(w) \cdot \tilde{{\mathbb T}}_c(|w|),
\end{eqnarray}
where
\begin{eqnarray}\label{r1-inf}
    R_\infty^{(1)}(w) = \frac{1}{2\pi}\, I_\nu(|w|)\, K_\nu(|w|)
\end{eqnarray}
is the `na\"{i}ve' limiting curve to be corrected (in a nonperturbative way) by the function $\tilde{{\mathbb T}}_c(|w|)= \tilde{{\mathbb T}}_c(|w|;n,\nu)$ proven to satisfy Eq.~(\ref{tc-exact}).
\noindent\newline\newline
Although Eq.~(\ref{r1-inf}) mirrors Eq.~(\ref{r-bulk}), it furnishes a limiting law which differs from Eq.~(\ref{r-large}); the difference derives from the fact that, in the Regime III, both the energies corresponding to the spectrum bulk and the parameter $\nu$ scale {\it linearly} in $n$. The latter necessitates the use of asymptotic expansions for the modified Bessel functions due to Olver, Lozier, Boisvert and Clark (2010):
\begin{eqnarray} \label{Iv-1}
    I_\nu(\nu z)&\simeq& \frac{1}{\sqrt{2\pi \nu}}\,\frac{e^{\nu\, \eta(z)}}{(1+z^2)^{1/4}} \sum_{k=0}^\infty \frac{U_k(p(z))}{\nu^k},\\
    \label{Iv-prime}
    I_\nu^\prime(\nu z)&\simeq& \frac{1}{\sqrt{2\pi \nu}}\,\frac{e^{\nu\, \eta(z)} (1+z^2)^{1/4}}{z} \sum_{k=0}^\infty \frac{V_k(p(z))}{\nu^k},
\end{eqnarray}
and
\begin{eqnarray}
    \label{Kv-1}
    K_\nu(\nu z)&\simeq& \sqrt{\frac{\pi}{2\nu}}\,\frac{e^{-\nu\, \eta(z)}}{(1+z^2)^{1/4}} \sum_{k=0}^\infty (-1)^k\frac{U_k(p(z))}{\nu^k},\\
    K_\nu^\prime(\nu z)&\simeq& - \sqrt{\frac{\pi}{2\nu}}\,\frac{e^{-\nu\, \eta(z)} (1+z^2)^{1/4}}{z} \sum_{k=0}^\infty (-1)^k\frac{V_k(p(z))}{\nu^k}.
\end{eqnarray}
Here, $z\in{\mathbb R}$, and the functions $\eta(z)$ and $p(z)$ are defined as
\begin{eqnarray}\label{eta-as}
    \eta(z) &=& \sqrt{1+z^2} + \log\,\frac{z}{1+ \sqrt{1+z^2}}, \\
    p(z) &=& \frac{1}{\sqrt{1+z^2}}.
\end{eqnarray}
Also, $U_k(p)$ and $V_k(p)$ are polynomials in $p$ of degree $3k$, given by $U_0(p)=V_0(p)=1$ and the recurrence
\begin{eqnarray}
    U_{k+1}(p) &=& \frac{1}{2}\, p^2(1-p^2)\, U_k^\prime(p) + \frac{1}{8} \int_0^p dt\, (1-5t^2)\, U_k(t), \\
    V_{k+1}(p) &=& U_{k+1}(p)-\frac{1}{2}\, p(1-p^2) \, U_k(p) - p^2(1-p^2) \,U_k^\prime(p).
\end{eqnarray}
\noindent\newline
These asymptotic expansions readily bring the large-$n$ formula
\begin{eqnarray}
\label{r-naive-2}
    R_\infty^{(1)}(w) = \frac{1}{4\pi} \, \frac{1}{\sqrt{|w^2|+\nu^2}}.
\end{eqnarray}
The slow decay at infinity exhibited by Eq.~(\ref{r-naive-2}) implies that this limiting law must break down at some `critical' point $|w|=|w_c|$. The mean density normalisation
\begin{eqnarray}
    2\pi \int_0^{|w_c|} d|w| \, |w| \times \frac{1}{4\pi} \, \frac{1}{\sqrt{|w|^2+\nu^2}} =n
\end{eqnarray}
suggests that
\begin{eqnarray}\label{crit-point-2}
|w_c| \simeq 2n \sqrt{q+1}.
\end{eqnarray}
The position of this break-down point is clearly seen in Fig. \ref{destruction}. \newline\newline\noindent
{\it Remark.}---Equations (\ref{r-naive-2}) and (\ref{crit-point-2}) can alternatively be derived within the two-dimensional Coulomb fluid approach discussed in Section \ref{Sec-2-3-1}. Indeed, making use of Eq.~(\ref{Kv-1}), we observe that the effective confinement potential in the Regime III is
\begin{eqnarray}\fl
    V_{\rm eff}(|w|)\Bigg|_{|w|\sim {\mathcal O}(\nu)} = \sqrt{|w|^2+\nu^2} - \nu \log \left(
        \nu + \sqrt{|w|^2+\nu^2}
    \right)\nonumber\\
    \fl \qquad \qquad\qquad\qquad\qquad + \frac{1}{4} \log (|w|^2+\nu^2)+ {\mathcal O}(|w|^{-1}).
\end{eqnarray}
Applying Eqs.~(\ref{dos-2d-cf}) and (\ref{dos-R-cf}), one reproduces, in the leading order in $1/\nu$, Eqs. (\ref{r-naive-2}) and (\ref{crit-point-2}), respectively.
$\;\;\;\blacksquare$
\newline\newline\noindent
To describe a vicinity of the critical point $|w|\sim |w_c|\simeq 2n\sqrt{q+1}$, we need to study the large-$|w|$ reduction of Eq.~(\ref{tc-exact}). Straightforward calculations based on Eqs.~(\ref{Iv-1}) and (\ref{Iv-prime}) yield:
\begin{eqnarray} \fl \label{t-eq-new}
    {\tilde{\mathbb T}}_c^{\prime\prime}(\varrho) + \frac{2\nu}{\varrho} \sqrt{1+ \frac{\varrho^2}{\nu^2}} \,{\tilde{\mathbb T}}_c^\prime(\varrho)
    = -\tilde{c}_{n,\nu} \, \varrho^{2n+\nu-2} \left(
        1+ \frac{\varrho^2}{\nu^2}
    \right)^{1/4} \exp\left[
        -\nu\, \eta \left(
            \frac{\varrho}{\nu}
        \right)
    \right].\nonumber\\
    {}
\end{eqnarray}
Here, the function $\eta(z)$ in the exponent is defined by Eq.~(\ref{eta-as}); the constant $\tilde{c}_{n,\nu}$ equals
\begin{eqnarray} \fl
    \tilde{c}_{n,\nu} = \frac{\sqrt{\pi \nu}}{2^{2n+\nu-5/2} \Gamma(n) \Gamma(n+\nu)} \simeq
    \sqrt{\frac{q(q+1)}{\pi}} \, \frac{e^{n(q+2)}}{(q+1)^{n(q+1)} (2n)^{n(q+2)-3/2}}.
\end{eqnarray}
Let us focus on a vicinity of the critical point by setting
\begin{eqnarray}
    |w|=\varrho = 2n\sqrt{q+1} +  2 n^\alpha t,\qquad t\sim {\mathcal O}(n^0),
\end{eqnarray}
where the exponent $\alpha$ will be determined later on. Introducing the new function
\begin{eqnarray}
    \tilde{\mathbb A}_c(t) = {\tilde{\mathbb T}}_c\left(\varrho= 2n\sqrt{q+1} +  2 n^\alpha t;n,nq\right),
\end{eqnarray}
one observes after somewhat lengthy but straightforward calculations that $\tilde{\mathbb A}_c(t)$ appears to have a well-defined large-$n$ limit provided $\alpha=1/2$. This limit is described by the differential equation
\begin{eqnarray}
    \tilde{\mathbb A}_c^\prime(t) = -\sqrt{\frac{2}{\pi (q+2)}}\, \exp\left(
        -\frac{2}{q+2}\,t^2
    \right)
\end{eqnarray}
supplemented by the initial condition $\tilde{\mathbb A}_c(-\infty)=1$. One has:
\begin{eqnarray}
    \tilde{\mathbb A}_c(t) = \frac{1}{2}\, {\rm erfc}\left(
          t \sqrt{\frac{2}{q+2}}    \right).
\end{eqnarray}
Equivalently,
\begin{eqnarray}
    \tilde{\mathbb T}_c(\varrho) = \frac{1}{2}\, {\rm erfc}\left(
          \frac{\varrho-2n \sqrt{q+1}}{\sqrt{2n(q+2)}}   \right).
\end{eqnarray}
Consequently, we conclude that the mean density of eigenlevels in the $\sqrt{n}$-vicinity of the critical point is given by the formula
\begin{eqnarray}
\label{mp-law-cp-2}
    R_{n,m}^{(1)}(w) = \frac{1}{8\pi}\,\frac{1}{\sqrt{|w|^2+n^2 q^2}}\, {\rm erfc}\left(
          \frac{|w|-2n \sqrt{q+1}}{\sqrt{2n(q+2)}}   \right).
\end{eqnarray}
Notice that formally performed $q\rightarrow 0$ limit reproduces Eq.~(\ref{crit-vic}) derived in Section \ref{Sec-2-3-1} for the case $n\gg 1$ and $\nu \sim {\mathcal O}(n^0)$.
\newline\newline\noindent
Equations (\ref{r-naive-2}) and (\ref{mp-law-cp-2}) represent the main result of this Subsection. Illustrated in Section~\ref{Sec-1-2-2}, they may naturally be considered as the complex-plane analogues of the Mar\u{c}enko-Pastur law [Eq.~(\ref{mp-1})]. It should be stressed that, contrary to the Mar\u{c}enko-Pastur law for real valued spectra, the mean density of complex eigenvalues does {\it not} develop a gap around the spectrum origin.

\section{Conclusions}
Driven by potential multidisciplinary applications in statistical analysis of {\it remote} complex systems characterised by {\it distinct} sets of stochastic time series, we have introduced a non-Hermitean extension of paradigmatic Wishart random matrices and presented a detailed study of their spectral properties in the simplest case of complex valued time series. In particular, we have determined (i) two alternative one-matrix-integral representations of the probability measure in the space of matrix entries [Eqs.~(\ref{mrs-01a}) and (\ref{mrs-01b})], (ii) the joint probability density function of all complex eigenvalues [Eq.~(\ref{msr-02})], (iii) the eigenvalue correlation functions of arbitrary (finite) order [Eqs.~(\ref{mrs-03a}) and (\ref{mrs-03b})], and also analysed (iv) the mean spectral density in various (large-$n$/large-$\nu$) scaling limits [Eqs.~(\ref{eq-ginue-sim}), (\ref{d-global}) and (\ref{mp-law-cp-2-list})].
\newline\newline\noindent
Interestingly (and also surprisingly for the authors), a family of closely related random matrix models has been introduced, and studied, in a totally different context of the QCD physics with a nonvanishing chemical potential (see, e.g., a review by Akemann (2007)). This observation (whatever expected or unexpected it may seem) provides one more evidence of the mysterious ubiquity of Random Matrix Theory appearing again and again in very distant fields of knowledge.

\section*{Acknowledgements}
The authors wish to thank G. Akemann for a clarifying correspondence and bringing many important references to our attention. This work was supported by the Israel Science Foundation through the grant No 414/08.

\newpage
\renewcommand{\appendixpagename}{\normalsize{Appendices}}
\addappheadtotoc
\appendixpage
\setcounter{section}{0}
\renewcommand{\thesection}{\Alph{section}}
\renewcommand{\theequation}{\thesection.\arabic{equation}}

\section{Three matrix integrals} \label{app-2}
In this Appendix, we give a detailed derivation of the three matrix integrals used in the main body of the paper. \newline\newline\noindent
{\bf Lemma 1.}~Let ${\boldsymbol Q}$ and ${\boldsymbol S}$ be $n\times n$ Hermitean and $n\times m$ rectangular complex matrices, respectively. Then, for $m\ge n$, it holds:
\begin{eqnarray}\fl
\label{b-01}
    \int_{{\mathbb C}^{n\times m}} D{\boldsymbol S} \, \delta\left(
        {\boldsymbol Q} - {\boldsymbol S}{\boldsymbol S}^\dagger
    \right) =
    \pi^{n(2m-n+1)/2} \frac{\prod_{j=1}^{m-n} \Gamma(j)}{\prod_{j=1}^{m} \Gamma(j)}\, ({\rm det} {\boldsymbol Q} )^{m-n} \, \Theta({\boldsymbol Q}).
\end{eqnarray}
\newline\newline\noindent
{\it Proof.}---Denoting the above integral as $I_{n,m}({\boldsymbol Q})$ and employing the matrix integral representation
\begin{eqnarray}
    \delta({\boldsymbol A}) = \frac{1}{2^n \pi^{n^2}} \int_{{\boldsymbol \sigma}^\dagger ={\boldsymbol \sigma}}
    D{\boldsymbol \sigma} \exp\left[
        i\,{\rm tr\,} \left(
            {\boldsymbol \sigma} {\boldsymbol A}
        \right)
    \right]
\end{eqnarray}
of the $\delta$-function of an $n\times n$ complex Hermitean matrix ${\boldsymbol A}$,
we set
\begin{eqnarray}
    {\boldsymbol A} = {\boldsymbol Q} - {\boldsymbol S}{\boldsymbol S}^\dagger
\end{eqnarray}
to write down:
\begin{eqnarray}\fl
\label{b-02}
    I_{n,m}({\boldsymbol Q})=\frac{1}{2^n \pi^{n^2}} \int_{{\boldsymbol \sigma}^\dagger ={\boldsymbol \sigma}}
    D{\boldsymbol \sigma} \,\exp\left[
    i\, {\rm tr\,} ({\boldsymbol \sigma} {\boldsymbol Q})\,
    \right]
    \int_{{\mathbb C}^{n\times m}} D{\boldsymbol S} \, \exp\left[
        - i {\rm tr\,} \left(
            {\boldsymbol S}^\dagger ({\boldsymbol \sigma} - i\eta) {\boldsymbol S}
        \right)
    \right]. \nonumber\\
    {}
\end{eqnarray}
Here and above, the flat measure $D{\boldsymbol \sigma}$ equals
\begin{eqnarray}\label{ds}
    D{\boldsymbol \sigma} = \prod_{j=1}^n d\sigma_{jj} \prod_{j>k=1}^n d\mathfrak{Re\,}\sigma_{jk}\, d\mathfrak{Im\,}\sigma_{jk},
\end{eqnarray}
and an infinitesimally small imaginary regulariser $-i\eta$ was introduced to ensure convergence of the Gaussian integral
\begin{eqnarray} \fl
    \int_{{\mathbb C}^{n\times m}} D{\boldsymbol S} \, \exp\left[
        - i {\rm tr\,} \left(
            {\boldsymbol S}^\dagger ({\boldsymbol \sigma} - i\eta) {\boldsymbol S}
        \right)
    \right] = (-i\pi)^{nm} {\rm det}^{-m} \left(
        {\boldsymbol \sigma} - i\eta {\mathds 1}_n
    \right).
\end{eqnarray}
Plugging this result back to Eq.~(\ref{b-02}), we derive:
\begin{eqnarray}\fl
\label{b-04}
    I_{n,m}({\boldsymbol Q})=\frac{(-i)^{nm}}{2^n \pi^{n(n-m)}} \int_{{\boldsymbol \sigma}^\dagger ={\boldsymbol \sigma}}
    D{\boldsymbol \sigma} \,\exp\left[
    i\, {\rm tr\,} ({\boldsymbol \sigma} {\boldsymbol Q})\,
    \right]\, {\rm det}^{-m} \left(
        {\boldsymbol \sigma} - i\eta {\mathds 1}_n
    \right).
\end{eqnarray}
This is the Ingham-Siegel integral of the second type (Ingham 1933, Siegel 1935, Fyodorov 2002) calculated in Lemma 2. For $m \ge n$, we make use of Eq.~(\ref{is-lemma}) to write down
\begin{eqnarray}
\label{b-04-04}\fl
    \int_{{\boldsymbol \sigma}^\dagger ={\boldsymbol \sigma}}
    D{\boldsymbol \sigma} \,\exp\left[
    i\, {\rm tr\,} ({\boldsymbol \sigma} {\boldsymbol Q})\,
    \right]\, {\rm det}^{-m} \left(
        {\boldsymbol \sigma} - i\eta {\mathds 1}_n
    \right) \nonumber \\
\fl \qquad\qquad= 2^n \frac{\prod_{j=1}^{m-n}\Gamma(j)}{\prod_{j=1}^m \Gamma(j)} \, \pi^{n(n+1)/2}
    i^{nm} \exp\left[
        -\eta\,{\rm tr\,} {\boldsymbol Q}
    \right] \, ({\rm det}{\boldsymbol Q})^{m-n}\Theta({\boldsymbol Q}).
\end{eqnarray}
Substituting Eq.~(\ref{b-04-04}) into Eq.~(\ref{b-04}) and letting $\eta$ tend to zero, we reproduce Eq.~(\ref{b-01}). End of proof. $\;\;\;\blacksquare$
\newline\newline\noindent
{\bf Lemma 2} \footnote{Although the proof below closely follows the one presented by Fyodorov (2002), see also Hua (1963), we have included a detailed derivation of Eq.~(\ref{is-lemma}) in order to prepare the reader for a proof of Lemma 3 where essentially the same idea is employed to tackle a more involved matrix integral.} {\bf (Ingham-Siegel Integral of the Second Type).}~Let  ${\boldsymbol Q}$ and ${\boldsymbol \sigma}$ be $n\times n$ Hermitean matrices, and $\mathfrak{Im}\, z >0$. Then, for $m \ge n$, it holds:
\begin{eqnarray}
\label{is-lemma}
\fl
    \int_{{\boldsymbol \sigma}^\dagger ={\boldsymbol \sigma}}
    D{\boldsymbol \sigma} \,\exp\left[
    i\, {\rm tr\,} ({\boldsymbol \sigma} {\boldsymbol Q})\,
    \right]\, {\rm det}^{-m} \left(
        {\boldsymbol \sigma} - z\, {\mathds 1}_n
    \right)\nonumber \\ \fl
    \qquad\qquad = 2^n \frac{\prod_{j=1}^{m-n}\Gamma(j)}{\prod_{j=1}^m \Gamma(j)} \, \pi^{n(n+1)/2}
    i^{nm} \exp\left[
        i z \,{\rm tr\,} {\boldsymbol Q}
    \right] \, ({\rm det}{\boldsymbol Q})^{m-n}\Theta({\boldsymbol Q}).
\end{eqnarray}
Here, $D{\boldsymbol \sigma}$ is given by Eq.~(\ref{ds}).
\newline\newline\noindent
{\it Proof.}---Due to the iterative nature of the forthcoming calculation, it is convenient to use the notation ${\boldsymbol Q}\equiv {\boldsymbol Q}_n$ and ${\boldsymbol \sigma}\equiv {\boldsymbol \sigma}_n$ that highlights the size of matrices in Eq.~(\ref{is-lemma}). Denoting the integral to be treated as $F_{n,m}({\boldsymbol Q}_n)$, we remark that it may only depend on the eigenvalues $(Q_1,\dots,Q_n)$ of ${\boldsymbol Q}_n$. The latter can be written in the decomposed form
\begin{eqnarray}
    {\boldsymbol Q}_n = \left(
                          \begin{array}{cc}
                            Q_1 & {\boldsymbol 0}^{\dagger}_{n-1} \\
                            {\boldsymbol 0}_{n-1} & {\boldsymbol Q}_{n-1} \\
                          \end{array}
                        \right),\quad {\boldsymbol Q}_{n-1} = {\rm diag} (Q_2,\dots,Q_{n}).
\end{eqnarray}
Accordingly, we represent the matrix ${\boldsymbol \sigma}_n$ in the block form
\begin{eqnarray}
    {\boldsymbol \sigma}_n = \left(
                               \begin{array}{cc}
                                 \sigma_{11} & {\boldsymbol u}^\dagger_{n-1} \\
                                 {\boldsymbol u}_{n-1} & {\boldsymbol \sigma}_{n-1} \\
                               \end{array}
                             \right),
\end{eqnarray}
where $\sigma_{11}$ is a real scalar, ${\boldsymbol \sigma}_{n-1}$ is a truncated matrix of the size $(n-1)\times (n-1)$, and
\begin{eqnarray}
{\boldsymbol u}_{n-1}^\dagger = (\bar{\sigma}_{12},\bar{\sigma}_{13},\dots,\bar{\sigma}_{1,n})
\end{eqnarray}
is a complex vector of the length $n-1$. In the above parameterisation, the integration measure in Eq.~(\ref{is-lemma}) factorises:
\begin{eqnarray}
    D{\boldsymbol\sigma}_n = d\sigma_{11}\, d\mathfrak{Re\,} {\boldsymbol u}_{n-1}\, d\mathfrak{Im\,} {\boldsymbol u}_{n-1}\, D{\boldsymbol \sigma}_{n-1}.
\end{eqnarray}
Owing to the identities
\begin{eqnarray}
    {\rm tr\,} ({\boldsymbol \sigma}_n{\boldsymbol Q}_n)
    =\Omega_{11} Q_1 + {\rm tr\,} ({\boldsymbol \sigma}_{n-1}{\boldsymbol Q}_{n-1}),
\end{eqnarray}
and
\begin{eqnarray}\fl
    {\rm det\,} ({\boldsymbol \sigma}_n - z\,{\mathds 1}_n)
    =
    {\rm det\,} \left(
                  \begin{array}{cc}
                    \sigma_{11}-z & {\boldsymbol u}_{n-1}^\dagger \\
                    {\boldsymbol u}_{n-1} & {\boldsymbol \sigma}_{n-1} - z\, {\mathds 1}_{n-1} \\
                  \end{array}
                \right)\nonumber\\ \fl
                \qquad
                = {\rm det\,} ({\boldsymbol \sigma}_{n-1} - z\,{\mathds 1}_{n-1})
                \left[
                    (\sigma_{11}-z) - {\boldsymbol u}_{n-1}^\dagger ({\boldsymbol \sigma}_{n-1} - z\,{\mathds 1}_{n-1})^{-1}
                    {\boldsymbol u}_{n-1}
                \right],
\end{eqnarray}
the sought integral $F_{n,m}({\boldsymbol Q}_n)$ can be decomposed:
\begin{eqnarray} \fl \label{int-00}
    F_{n,m}({\boldsymbol Q}_n) =
    \int_{{\boldsymbol \sigma}_{n-1}^\dagger ={\boldsymbol \sigma}_{n-1}}
    D{\boldsymbol \sigma}_{n-1} \,\exp\left[
    i\, {\rm tr\,} ({\boldsymbol \sigma}_{n-1} {\boldsymbol Q}_{n-1})\,
    \right]\, {\rm det}^{-m} \left(
        {\boldsymbol \sigma}_{n-1} - z\, {\mathds 1}_{n-1}
    \right)\nonumber \\
    \fl \qquad
    \times \int_{{\mathbb R}^{n-1}} d \mathfrak{Re\,} {\boldsymbol u}_{n-1} \int_{{\mathbb R}^{n-1}} d \mathfrak{Im\,} {\boldsymbol u}_{n-1}
    \nonumber\\
    \fl \qquad   \times   \int_{\mathbb R} d\sigma_{11} \exp[i \sigma_{11} Q_1]
    \left[
                    (\sigma_{11}-z) - {\boldsymbol u}_{n-1}^\dagger ({\boldsymbol \sigma}_{n-1} - z\,{\mathds 1}_{n-1})^{-1}
                    {\boldsymbol u}_{n-1}
                \right]^{-m}.
\end{eqnarray}
The integral over the scalar $\sigma_{11}$ can be calculated by the residue theorem which, for $\mathfrak{Im\,}z > 0$, yields
\begin{eqnarray} \fl
    \int_{\mathbb R} d\sigma_{11} \exp[i \sigma_{11} Q_1]
    \left[
                    (\sigma_{11}-z) - {\boldsymbol u}_{n-1}^\dagger ({\boldsymbol \sigma}_{n-1} - z\,{\mathds 1}_{n-1})^{-1}
                    {\boldsymbol u}_{n-1}
                \right]^{-m}\nonumber \\
                \fl \qquad = 2\pi \frac{i^m}{\Gamma(m)}
                \Theta(Q_1)\, Q_1^{m-1}\exp\left[
                    i Q_1 \left(
                      z+ u_{n-1}^\dagger ({\boldsymbol \sigma}_{n-1} - z\,{\mathds 1}_{n-1})^{-1}
                    {\boldsymbol u}_{n-1}
                    \right)
                    \right].
\end{eqnarray}
Now we are left with the Gaussian integral whose calculation is straightforward:
\begin{eqnarray}\fl
   \int_{{\mathbb R}^{n-1}} d \mathfrak{Re\,} {\boldsymbol u}_{n-1} \int_{{\mathbb R}^{n-1}} d \mathfrak{Im\,} {\boldsymbol u}_{n-1}
   \exp\left[
                    i Q_1 \left(
                      z+ {\boldsymbol u}_{n-1}^\dagger ({\boldsymbol \sigma}_{n-1} - z\,{\mathds 1}_{n-1})^{-1}
                    {\boldsymbol u}_{n-1}
                    \right)
                    \right] \nonumber\\
                    \fl \qquad
                    = \left(
                    \frac{i\pi}{Q_1}
                    \right)^{m-1}
           {\rm det} \left(
        {\boldsymbol \sigma}_{n-1} - z\, {\mathds 1}_{n-1}
    \right).
\end{eqnarray}
Plugging this result back into Eq.~(\ref{int-00}), we observe the recurrence relation:
\begin{eqnarray} \label{chain-first}
    \frac{F_{n,m}({\boldsymbol Q}_n)}{F_{n-1,m-1}({\boldsymbol Q}_{n-1})} = \frac{2}{\Gamma(m)} \pi^n i^{n+m-1} \Theta(Q_1) \, Q_1^{m-n} \exp[i Q_1 z].
\end{eqnarray}
Assuming that $m \ge n$, we iterate the recurrence relation $n-1$ times to obtain the chain of equations:
\begin{eqnarray}
    \hspace{-1cm}\frac{F_{n-1,m-1}({\boldsymbol Q}_{n-1})}{F_{n-2,m-2}({\boldsymbol Q}_{n-2})} &=& \frac{2}{\Gamma(m-1)} \pi^{n-1} i^{n+m-3} \Theta(Q_2) \, Q_2^{m-n} \exp[i Q_2 z],\nonumber\\
    \hspace{-1cm}\frac{F_{n-2,m-2}({\boldsymbol Q}_{n-2})}{F_{n-3,m-3}({\boldsymbol Q}_{n-3})} &=& \frac{2}{\Gamma(m-2)} \pi^{n-2} i^{n+m-5} \Theta(Q_3) \, Q_3^{m-n} \exp[i Q_3 z], \nonumber\\
    \hspace{-1cm}&\dots& \nonumber\\
    \hspace{-1cm}\frac{F_{2,m-n+2}({\boldsymbol Q}_{2})}{F_{1,m-n+1}({\boldsymbol Q}_{1})} &=& \frac{2}{\Gamma(m-n+2)} \pi^{2} i^{m-n+3} \Theta(Q_{n-1}) \, Q_{n-1}^{m-n} \exp[i Q_{n-1} z],
    \nonumber\\
    {}
\end{eqnarray}
where
\begin{eqnarray} \label{chain-last}
    \hspace{-2cm}F_{1,m-n+1}({\boldsymbol Q}_{1}) &=& \int_{\mathbb R} d\sigma_{nn} \exp[i \sigma_{nn} Q_{n}]\, (\sigma_{nn}-z)^{m-n+1}
    \nonumber\\
    &=& \frac{2}{\Gamma(m-n+1)} \pi^{1} i^{m-n+1} \Theta(Q_{n}) \, Q_{n}^{m-n} \exp[i Q_{n} z].
\end{eqnarray}
Combining Eqs.~(\ref{chain-first}) -- (\ref{chain-last}) and restoring the initial notation, we derive:
\begin{eqnarray}\fl
    F_{n,m}({\boldsymbol Q}) = \frac{2^n}{\prod_{j=m-n+1}^m \Gamma(j)}\, \pi^{n(n+1)/2} i^{nm} \, \exp[i z\, {\rm tr\,}{\boldsymbol Q}]\, ({\rm det\,} {\boldsymbol Q})^{m-n} \Theta({\boldsymbol Q}).
\end{eqnarray}
This is equivalent to the r.h.s. of Eq.~(\ref{is-lemma}). End of proof.  $\;\;\;\blacksquare$
\newline\newline\noindent
{\bf Lemma 3.}~Let  ${\boldsymbol \psi}$ be an $n\times n$ complex valued lower triangular matrix
\begin{eqnarray}
    ({\boldsymbol \psi})_{jk} = \left\{
                               \begin{array}{cc}
                                 0, & \hbox{$j<k$}, \\
                                 \psi_{jk}\in {\mathbb C}, & \hbox{$j\ge k$,}
                               \end{array}
                             \right.
\end{eqnarray}
and ${\boldsymbol w}$ be an $n\times n$ diagonal matrix ${\boldsymbol w}=(w_1,\dots,w_n)$. It holds that
\begin{eqnarray}\fl \label{lem-03-01}
        \int_{{\boldsymbol \psi} \in {\mathbb C}^{n(n+1)/2}} D{\boldsymbol \psi}\,
    \exp\left[
         \frac{i}{2}\, {\rm tr} \left(
            {\boldsymbol \psi} {\boldsymbol w}^\dagger + {\boldsymbol \psi}^\dagger {\boldsymbol w}
        \right)
    \right] {\rm det}^{-m} \left(
    {\mathds 1}_n + {\boldsymbol \psi}{\boldsymbol \psi}^\dagger
    \right) \nonumber\\
    = \frac{\pi^{n(n+1)/2}}{2^{n(m-n-1)}} \frac{\prod_{j=1}^{m-n} \Gamma(j)}{\prod_{j=1}^m \Gamma(j)}
    \,\prod_{j=1}^n |w_j|^{m-n} \, K_{m-n}(|w_j|).
\end{eqnarray}
Here, $K_n(w)$ is the modified Bessel function of the second kind.
\newline\newline\noindent
{\it Proof.}---Due to the iterative nature of the forthcoming calculation, it is convenient to use the notation ${\boldsymbol \psi}\equiv {\boldsymbol \psi}_n$ and ${\boldsymbol w}\equiv {\boldsymbol w}_n$ that highlights the size of matrices in Eq.~(\ref{lem-03-01}). Denoting the integral to be treated as $I_{n,m}({\boldsymbol w}_n)$, we proceed with its calculation by applying the method used in Lemma 2. Specifically, we decompose the $n\times n$ matrix ${\boldsymbol \psi}_n$ as
\begin{eqnarray}
    {\boldsymbol \psi}_n = \left(
                             \begin{array}{cc}
                               {\boldsymbol \psi}_{n-1} & {\boldsymbol 0}_{n-1} \\
                               {\boldsymbol u}_{n-1} & \psi_{nn} \\
                             \end{array}
                           \right),
\end{eqnarray}
where $\psi_{nn}$ is a complex scalar, ${\boldsymbol \psi}_{n-1}$ is a truncated lower triangular matrix of the size $(n-1)\times (n-1)$, and
\begin{eqnarray}
    {\boldsymbol u}_{n-1}^\dagger = (\bar{\psi}_{n,1}, \bar{\psi}_{n,2},\cdots, \bar{\psi}_{n,n-1})
\end{eqnarray}
is a complex vector of the length $n-1$. In the above parameterisation, the integration measure $D{\boldsymbol \psi}\equiv D{\boldsymbol \psi}_n$
in Eq.~(\ref{lem-03-01}) factorises:
\begin{eqnarray}
\label{int-meas}
    D{\boldsymbol \psi}_n =  d\mathfrak{Re\,}\psi_{nn}\,d\mathfrak{Im\,}\psi_{nn}\, d\mathfrak{Re\,}{\boldsymbol u}_{n-1}
    d\mathfrak{Im\,}{\boldsymbol u}_{n-1}\, D{\boldsymbol \psi}_{n-1}.
\end{eqnarray}
Owing to the identities
\begin{eqnarray}\fl
\label{tr-id}
     {\rm tr\,}({\boldsymbol \psi}_n \boldsymbol{w}_n^\dagger +
        {\boldsymbol \psi}_n^\dagger \boldsymbol{w}_n) =
        \psi_{nn} \bar{w}_n + \bar{\psi}_{nn} w_n +
        {\rm tr\,}(
        {\boldsymbol \psi}_{n-1} \boldsymbol{w}_{n-1}^\dagger +
        {\boldsymbol \psi}_{n-1}^\dagger \boldsymbol{w}_{n-1})
\end{eqnarray}
and
\begin{eqnarray}\fl
    {\rm det\,}({\mathds 1}_n + {\boldsymbol \psi}_{n} {\boldsymbol \psi}_{n}^\dagger) =
    {\rm det\,} \left(
                             \begin{array}{cc}
                               {\mathds 1}_{n-1} + {\boldsymbol \psi}_{n-1}{\boldsymbol \psi}_{n-1}^\dagger & {\boldsymbol \psi}_{n-1} {\boldsymbol u}_{n-1} \\
                               {\boldsymbol u}_{n-1}^\dagger {\boldsymbol \psi}_{n-1}^\dagger & 1 + {\boldsymbol u}_{n-1}^\dagger {\boldsymbol u}_{n-1} + |\psi_{nn}|^2\\
                             \end{array}
                             \right) \nonumber\\
   \fl \qquad \quad=
    {\rm det\,}({\mathds 1}_{n-1} + {\boldsymbol \psi}_{n-1} {\boldsymbol \psi}_{n-1}^\dagger) \nonumber\\
    \fl \qquad \quad \times
    \left[
        1+ |\psi_{nn}|^2 + {\boldsymbol u}_{n-1}^\dagger
        \bigg(
        {\mathds 1}_{n-1} -
        {\boldsymbol \psi}_{n-1}^\dagger
        ({\mathds 1}_{n-1} + {\boldsymbol \psi}_{n-1} {\boldsymbol \psi}_{n-1}^\dagger)^{-1}
        {\boldsymbol \psi}_{n-1}
        \bigg) \, {\boldsymbol u}_{n-1}
    \right], \nonumber\\
{}
\end{eqnarray}
the integral $I_{n,m}({\boldsymbol w}_n)$ can be decomposed as follows:
\begin{eqnarray}
\label{Inp-02}\fl
    I_{n,m}({\boldsymbol w}_n) =
    \int_{{\boldsymbol \psi}_{n-1} \in{\mathbb C}^{n(n-1)/2}} D{\boldsymbol \psi}_{n-1}
        \exp\left[\frac{i}{2}\, {\rm tr}\left(
        {\boldsymbol \psi}_{n-1} \boldsymbol{w}_{n-1}^\dagger +
        {\boldsymbol \psi}_{n-1}^\dagger \boldsymbol{w}_{n-1}\right)
    \right] \nonumber\\
    \fl \qquad \quad \times \;
    {\rm det}^{-m} ({\mathds 1}_{n-1}+ {\boldsymbol \psi}_{n-1} {\boldsymbol \psi}_{n-1}^\dagger)
    \,
\int_{\mathbb R} d\mathfrak{Re\,}\psi_{nn}\,\int_{\mathbb R} d\mathfrak{Im\,}\psi_{nn}\,
    \nonumber\\
\fl \qquad\quad \times \;\exp\left[\frac{i}{2}\left(\psi_{nn} \bar{w}_n + \bar{\psi}_{nn} w_n\right)\right]\,
\int_{{\mathbb R}^{n-1}} \, d\mathfrak{Re\,} {\boldsymbol u}_{n-1} \int_{{\mathbb R}^{n-1}} \, d\mathfrak{Im\,} {\boldsymbol u}_{n-1}
    \nonumber\\
    \fl
    \qquad \quad \times
    \; \left[
            1+ |\psi_{nn}|^2 + {\boldsymbol u}_{n-1}^\dagger
        \bigg(
        {\mathds 1}_{n-1} -
        {\boldsymbol \psi}_{n-1}^\dagger
        ({\mathds 1}_{n-1} + {\boldsymbol \psi}_{n-1} {\boldsymbol \psi}_{n-1}^\dagger)^{-1}
        {\boldsymbol \psi}_{n-1}
        \bigg) \, {\boldsymbol u}_{n-1}
    \right]^{-m}. \nonumber\\
{}
\end{eqnarray}
(i) Let us first calculate the inner integral
\begin{eqnarray}\fl\label{L-int}
    L({\boldsymbol \psi}_{n-1},{\boldsymbol \psi}_{n-1}^\dagger, |\psi_{nn}|) = \int_{{\mathbb R}^{n-1}} \, d\mathfrak{Re\,} {\boldsymbol u}_{n-1} \int_{{\mathbb R}^{n-1}} \, d\mathfrak{Im\,} {\boldsymbol u}_{n-1}
    \nonumber\\
    \fl
    \qquad \quad \times
    \; \left[
            1+ |\psi_{nn}|^2 + {\boldsymbol u}_{n-1}^\dagger
        \bigg(
        {\mathds 1}_{n-1} -
        {\boldsymbol \psi}_{n-1}^\dagger
        ({\mathds 1}_{n-1} + {\boldsymbol \psi}_{n-1} {\boldsymbol \psi}_{n-1}^\dagger)^{-1}
        {\boldsymbol \psi}_{n-1}
        \bigg) \, {\boldsymbol u}_{n-1}
    \right]^{-m}. \nonumber\\
{}
\end{eqnarray}
To this end, we introduce a new (vector) integration variable
\begin{eqnarray}
    {\boldsymbol v}_{n-1} = {\boldsymbol M}^{1/2}_{n-1} {\boldsymbol u}_{n-1},
\end{eqnarray}
where an $(n-1)\times (n-1)$ Hermitean matrix ${\boldsymbol M}_{n-1}$ is given by
\begin{eqnarray} \fl
        {\boldsymbol M}_{n-1} = \frac{1}{1+|\psi_{nn}|^2} \left[
        {\mathds 1}_{n-1} -
        {\boldsymbol \psi}_{n-1}^\dagger
        ({\mathds 1}_{n-1} + {\boldsymbol \psi}_{n-1} {\boldsymbol \psi}_{n-1}^\dagger)^{-1}
        {\boldsymbol \psi}_{n-1}
        \right].
\end{eqnarray}
Since the flat measure $d\mathfrak{Re\,} {\boldsymbol u}_{n-1}\,d\mathfrak{Im\,} {\boldsymbol u}_{n-1}$ transforms as follows,
\begin{eqnarray} \fl
    d\mathfrak{Re\,} {\boldsymbol u}_{n-1}\,d\mathfrak{Im\,} {\boldsymbol u}_{n-1}
    =\frac{1}{{\rm det} {\boldsymbol M}_{n-1}}\, d\mathfrak{Re\,} {\boldsymbol v}_{n-1}\,d\mathfrak{Im\,} {\boldsymbol v}_{n-1} \nonumber\\
    \fl \qquad\qquad
    = (1+|\psi_{nn}|)^{n-1} \, {\rm det} ({\mathds 1}_{n-1}+ {\boldsymbol \psi}_{n-1} {\boldsymbol \psi}_{n-1}^\dagger)
\, d\mathfrak{Re\,} {\boldsymbol v}_{n-1}\,d\mathfrak{Im\,} {\boldsymbol v}_{n-1},
\end{eqnarray}
the integral Eq.~(\ref{L-int}) boils down to
\begin{eqnarray}\fl
    L({\boldsymbol \psi}_{n-1},{\boldsymbol \psi}_{n-1}^\dagger, |\psi_{nn}|) =
    (1+|\psi_{nn}|^2)^{n-m-1} \, {\rm det} ({\mathds 1}_{n-1}+ {\boldsymbol \psi}_{n-1} {\boldsymbol \psi}_{n-1}^\dagger) \nonumber\\
\qquad \times\,
        \int_{{\mathbb R}^{n-1}} \, d\mathfrak{Re\,} {\boldsymbol v}_{n-1} \int_{{\mathbb R}^{n-1}} \, d\mathfrak{Im\,} {\boldsymbol v}_{n-1}
    \,
    \left(
            1+ {\boldsymbol v}_{n-1}^\dagger\, {\boldsymbol v}_{n-1}
    \right)^{-m}
\end{eqnarray}
As the remaining integral is straightforward to calculate,
\begin{eqnarray}
    \fl
    \int_{{\mathbb R}^{n-1}} \, d\mathfrak{Re\,} {\boldsymbol v}_{n-1} \int_{{\mathbb R}^{n-1}} \, d\mathfrak{Im\,} {\boldsymbol v}_{n-1}
    \,
    \left(
            1+ {\boldsymbol v}_{n-1}^\dagger\, {\boldsymbol v}_{n-1}
    \right)^{-m} = \pi^{n-1} \frac{\Gamma(m-n+1)}{\Gamma(m)},
\end{eqnarray}
we conclude that
\begin{eqnarray}\fl\label{L-int-answer}
    L({\boldsymbol \psi}_{n-1},{\boldsymbol \psi}_{n-1}^\dagger, |\psi_{nn}|) =
    \pi^{n-1} \frac{\Gamma(m-n+1)}{\Gamma(m)}  \,
    (1+|\psi_{nn}|^2)^{n-m-1} \, {\rm det} ({\mathds 1}_{n-1}+ {\boldsymbol \psi}_{n-1} {\boldsymbol \psi}_{n-1}^\dagger).
\nonumber \\ {}
\end{eqnarray}
Now, the integral $I_{n,m}({\boldsymbol w}_n)$ simplifies to
\begin{eqnarray} \fl
    I_{n,m}({\boldsymbol w}_n) = \pi^{n-1} \frac{\Gamma(m-n+1)}{\Gamma(m)} \, \int_{{\boldsymbol \psi}_{n-1} \in {\mathbb C}^{n(n-1)/2}} D{\boldsymbol \psi}_{n-1}\, \nonumber\\
    \fl \qquad \times \,
    \exp\left[
         \frac{i}{2}\, {\rm tr} \left(
            {\boldsymbol \psi}_{n-1} {\boldsymbol w}^\dagger_{n-1} + {\boldsymbol \psi}_{n-1}^\dagger {\boldsymbol w}_{n-1}
        \right)
    \right] \, {\rm det}^{-(m-1)} \left(
    {\mathds 1}_{n-1} + {\boldsymbol \psi}_{n-1}{\boldsymbol \psi}^\dagger_{n-1}
    \right) \nonumber\\
    \fl\qquad \times \,
\left(
        \int_{{\mathbb R}} d\mathfrak{Re\,}\psi_{nn}\, \int_{{\mathbb R}} d\mathfrak{Im\,}\psi_{nn}
        \,
        \exp\left[\frac{i}{2}\left(\psi_{nn} \bar{w}_n + \bar{\psi}_{nn} w_n\right)\right]\,
          (1+|\psi_{nn}|^2)^{n-m-1}
\right). \nonumber\\
{}
\end{eqnarray}
\newline\noindent
(ii) The above representation is remarkable as the integral over the reduced matrix ${\boldsymbol \psi}_{n-1}$ can be recognised to be $I_{n-1,m-1}({\boldsymbol w}_{n-1})$. This observation, considered in conjunction with the result
\begin{eqnarray} \fl
    \int_{{\mathbb R}} d\mathfrak{Re\,}\psi_{nn}\, \int_{{\mathbb R}} d\mathfrak{Im\,}\psi_{nn}
        \,
        \exp\left[\frac{i}{2}\left(\psi_{nn} \bar{w}_n + \bar{\psi}_{nn} w_n\right)\right]\,
          (1+|\psi_{nn}|^2)^{n-m-1} \nonumber\\
        \fl \qquad \qquad \qquad \qquad \qquad  = \frac{2\pi}{2^{m-n} \Gamma(m-n+1)}
\, |w_n|^{m-n} K_{m-n}(|w_n|),
\end{eqnarray}
yields the recursion relation:
\begin{eqnarray} \label{c1}
    \frac{I_{n,m}({\boldsymbol w}_n)}{I_{n-1,m-1}({\boldsymbol w}_{n-1})} = \frac{(2\pi)^n}{2^{m-1} \Gamma(m)}
\, |w_n|^{m-n} K_{m-n}(|w_n|).
\end{eqnarray}
Assuming that $m\ge n$, we iterate it $(n-1)$ times to obtain the chain of equations:
\begin{eqnarray}\fl\label{c2}
    \qquad
        \frac{I_{n-1,m-1}({\boldsymbol w}_{n-1})}{I_{n-2,m-2}({\boldsymbol w}_{n-2})} = \frac{(2\pi)^{n-1}}{2^{m-2} \Gamma(m-1)}
\, |w_{n-1}|^{m-n} K_{m-n}(|w_{n-1}|), \nonumber\\
\fl
    \qquad
        \frac{I_{n-2,m-2}({\boldsymbol w}_{n-2})}{I_{n-3,m-3}({\boldsymbol w}_{n-3})} = \frac{(2\pi)^{n-2}}{2^{m-3} \Gamma(m-2)}
\, |w_{n-2}|^{m-n} K_{m-n}(|w_{n-2}|), \nonumber\\
\fl \qquad \qquad\qquad\qquad \dots \nonumber\\
\fl
    \qquad
        \frac{I_{2,m-n+2}({\boldsymbol w}_{2})}{I_{1,m-n+1}({\boldsymbol w}_{1})} = \frac{(2\pi)^{2}}{2^{m-n+1} \Gamma(m-n+2)}
\, |w_{2}|^{m-n} K_{m-n}(|w_{2}|),
\end{eqnarray}
where
\begin{eqnarray}\fl\label{c3}
    I_{1,m-n+1}({\boldsymbol w}_{1}) =
    \int_{{\mathbb R}} d\mathfrak{Re\,}\psi_{11}\, \int_{{\mathbb R}} d\mathfrak{Im\,}\psi_{11}
        \,
        \exp\left[\frac{i}{2}\left(\psi_{11} \bar{w}_1 + \bar{\psi}_{11} w_1\right)\right]\,
          (1+|\psi_{11}|^2)^{n-m-1} \nonumber\\
        \fl \qquad \qquad \qquad \qquad \qquad  = \frac{(2\pi)^1}{2^{m-n} \Gamma(m-n+1)}
\, |w_1|^{m-n} K_{m-n}(|w_1|).
\end{eqnarray}
Combining Eqs.~(\ref{c1}) -- (\ref{c3}) and restoring the initial notation, we derive:
\begin{eqnarray}\fl\qquad
    I_{n,m}({\boldsymbol w}) = \frac{\pi^{n(n+1)}}{2^{n(m-n-1)}}
    \frac{1}{\prod_{j=m-n+1}^m \Gamma(j)}
    \,\prod_{j=1}^n |w_j|^{m-n} K_{m-n}(|w_j|).
\end{eqnarray}
This is equivalent to the r.h.s. of Eq.~(\ref{lem-03-01}). End of proof. $\;\;\;\blacksquare$

\section{Asymptotic behaviour of the Bessel function $K_\nu(x \sqrt{\nu})$ as $\nu \rightarrow \infty$}\label{app-3}
{\bf Lemma 4.} Let $K_\nu(x)$ be the modified Bessel function of the second kind. The main term of the asymptotic expansion of $K_\nu(x \sqrt{\nu})$ as
$\nu \rightarrow \infty$ equals
\begin{eqnarray}
\label{lem-c-00}
    K_\nu(x\sqrt{\nu}) \simeq  \left(
        \frac{4\nu}{x^2}
    \right)^{\nu/2}  \sqrt{\frac{\pi}{2 \nu}}\, e^{-\nu}\, e^{-x^2/4},
\end{eqnarray}
where $x>0$.
\newline\newline\noindent
{\it Proof.}---The easiest way to see that Eq.~(\ref{lem-c-00}) holds true is to start with Basset's integral \footnote{See Section 7.8 of Olver (1974).}
\begin{eqnarray}
\label{lem-c-01}
    K_\nu(xz) = \frac{1}{\sqrt{\pi}} \left(
        \frac{2z}{x}
    \right)^\nu \Gamma\left(
    \nu+\frac{1}{2}
    \right) \int_0^\infty dt\, \frac{\cos (xt)}{(t^2+z^2)^{\nu+1/2}},
\end{eqnarray}
where $\mathfrak{Re\,} \nu > -1/2$, $x>0$, and ${\rm arg}(z) < \pi/2$; the branch of $(t^2 + z^2)^{\nu+1/2}$ is continuous and asymptotic to the principal value
of $t^{2\nu+1}$ as $t\rightarrow +\infty$. Setting $z=\sqrt{\nu}$, we observe the large-$\nu$ relation
\begin{eqnarray}
    \frac{1}{(t^2+\nu)^{\nu+1/2}}\Bigg|_{\nu \gg 1} \simeq \frac{1}{\nu^{\nu+1/2}} \, e^{-t^2}.
\end{eqnarray}
Plugging it back to Eq.~(\ref{lem-c-01}) and making use of the leading order Stirling formula
\begin{eqnarray}
    \Gamma\left( \nu+ \frac{1}{2}\right)\Bigg|_{\nu \gg 1} \simeq \sqrt{2\pi} \, \nu^{\nu} e^{-\nu},
\end{eqnarray}
we derive:
\begin{eqnarray}
\label{lem-c-02}
    K_\nu(x\sqrt{\nu}) \simeq \left(
        \frac{4\nu}{x^2}
    \right)^{\nu/2}  \sqrt{\frac{2}{\nu}}\, e^{-\nu} \int_0^\infty dt\, \cos (xt)\, e^{-t^2}.
\end{eqnarray}
Performing the integral, we reproduce Eq.~(\ref{lem-c-00}). End of proof.  $\;\;\;\blacksquare$

\smallskip
\section*{References}
\fancyhead{} \fancyhead[RE,LO]{References}
\fancyhead[LE,RO]{\thepage}
\addcontentsline{toc}{section}{\protect\enlargethispage*{100pt}References}
\begin{harvard}

\item[] Akemann G 2007
        Matrix models and QCD with chemical potential
        {\it Int. J. Mod. Phys.} A {\bf 22} 1077 -- 1122

\item[] Akemann G, Bloch J, Shifrin L and Wettig T 2008
        Individual complex Dirac eigenvalue distributions from random matrix theory and comparison to quenched lattice QCD with a quark
        chemical potential
        \PRL {\bf 100} 032002-1 -- 032002-4

\item[] Akemann G and Kanzieper E 2007
        Integrable structure of Ginibre's ensemble of real random matrices and a Pfaffian integration theorem
        {\it J. Stat. Phys.} {\bf 129} 1159 -- 1231

\item[] Akemann G, Osborn J C, Splittorff K and Verbaarschot J J M 2005
        Unquenched QCD Dirac operator spectra at nonzero baryon chemical potential
        {\it Nucl. Phys.} B {\bf 712} 287 -- 324

\item[] Akemann G, Phillips M J and Shifrin L 2009
        Gap probabilities in non-Hermitian random matrix theory
        {\it J. Math. Phys.} {\bf 50} 063504-1 -- 063504-32

\item[] Andr\'eief C 1883
        Note sur une relation les int´egrales d´efinies des produits des fonctions
        {\it M\'em. de la Soc. Sci.} {\bf 2} 1 -- 14

\item[] Bai Z D 1997
        Circular law
        {\it Ann. Probab.} {\bf 25} 494 -- 529

\item[] Barth\'elemy M, Gondran B and Guichard E 2002
        Large scale cross-correlations in Internet traffic
        \PR E {\bf 66} 056110-1 -- 056110-7

\item[] Basu G, Ray K and Panigrahi P K 2010
        Random matrix route to image denoising
        \texttt{arXiv:1004.1356}

\item[] Biely C and Thurner S 2008
        Random matrix ensembles of time-lagged correlation matrices: Derivation of eigenvalue spectra and analysis of financial time-series
        {\it Quant. Finance} {\bf 8} 705 -- 722

\item[] Bouchaud J-P and Potters M 2009
        Financial applications of random matrix theory: A short review
        \texttt{arXiv:0910.1205}

\item[] Burda Z, Janik R A and Waclaw B 2010
        Spectrum of the product of independent random Gaussian matrices
        {\it Phys. Rev. E} {\bf 81} 041132-1 -- 041132-12

\item[] Chau L-L and Zaboronsky O 1998
        On the structure of correlation functions in the normal matrix model
        {\it Commun. Math. Phys.} {\bf 196} 203 -- 247

\item[] de Bruijn N G 1955
        On some multiple integrals involving determinants
        {\it J. Indian Math. Soc.} {\bf 19} 133 -- 151

\item[] Dyson F J 1971
        Distribution of eigenvalues for a class of real symmetric matrices
        {\it Rev. Mex. Fis.} {\bf 20} 231 -- 237

\item[] Fyodorov Y V 2002
        Negative moments of characteristic polynomials of random matrices: Ingham-Siegel integral as an alternative to Hubbard-Stratonovich transformation
        {\it Nucl. Phys.} B {\bf 621\,[PM]} 643 -- 674

\item[] Ginibre J 1965
        Statistical ensembles of complex, quaternion, and real
        matrices
        \JMP {\bf 6} 440 -- 449

\item[] Girko V L 1984
        Circular law
        {\it Theory Probab. Appl.} {\bf 29} 694 -- 706

\item[] Girko V L 1986
        Elliptic law
        {\it Theory Probab. Appl.} {\bf 30} 677 -- 690

\item[] Hua L K 1963
        {\it Harmonic Analysis of Functions of Several Complex Variables in the Classical Domains}
        (Providence: American Mathematical Society)

\item[] Ingham A E 1933
        An integral which occurs in statistics
        {\it Proc. Cambridge Philos. Soc.} {\bf 29} 271 -- 276

\item[] Janik R A and Nowak M A 2003
        Wishart and anti-Wishart random matrices
        {\it J. Phys. A: Math. Gen.} {\bf 36} 3629 -- 3637

\item[] Kanzieper E 2005
        Exact replica treatment of non-Hermitean complex random matrices, in:
        O. Kovras (ed.) {\it Frontiers in Field Theory} (New York: Nova Science Publishers)

\item[] Kwapie\'n J, Dro\.zd\.z S, Liu L C and Ioannides A A 1998
        Cooperative dynamics in auditory brain response
        \PR E {\bf 58} 6359 -- 6367

\item[] Kwapie\'{n} J, Dro\.{z}d\.{z} S, and Ioannides A A 2000
        Temporal correlations versus noise in the correlation matrix
        formalism: An example of the brain auditory response
        \PR E {\bf 62} 5557 -- 5564

\item[] Kwapie\'{n} J, Dro\.{z}d\.{z} S, G\'{o}rski A Z, and
        O\'{s}wi\c{e}cimka P
        2006
        Asymmetric matrices in an analysis of financial correlations
        {\it Acta Phys. Polonica} {\bf B37} 3039 -- 3048

\item[] Laloux L, Cizeau P, Bouchaud J-P and Potters M 1999
        Noise dressing of financial correlation matrices
        \PRL {\bf 83} 1467 -- 1470

\item[] Mar\u{c}enko V A and Pastur L A 1967
        Distribution of eigenvalues for some sets of random matrices
        {\it Math. USSR Sbornik} {\bf 1} 457 -- 483

\item[] Mehta M.L. 1976
        A note on certain multiple integrals
        {\it J. Math. Phys.} {\bf 17} 2198 -- 2202

\item[] Mehta M L 2004
        {\it Random Matrices} (Amsterdam: Elsevier)

\item[] Muirhead R J 1982
        {\it Aspects of Multivariate Statistical Theory} (New York: John
        Wiley \& Sons)

\item[] Olver F W J 1974
        {\it Asymptotics and Special Functions} (New York: Academic Press)

\item[] Olver F W J, Lozier D W, Boisvert R F and Clark C W 2010
        {\it NIST Handbook of Mathematical Functions} (Cambridge: Cambridge University Press)

\item[] Osborn J C 2004
        Universal results from an alternate random-matrix model for QCD with a baryon chemical potential
        \PRL {\bf 93} 222001-1 -- 222001--4

\item[] Plerou V, Gopikrishnan P, Rosenow B, Amaral L A N and Stanley H E 1999
        Universal and non-universal properties of cross-correlations in financial time series
        \PRL {\bf 83} 1471 -- 1474

\item[] Plerou V, Gopikrishnan P, Rosenow B, Amaral L A N and Stanley H E 2000
        A random matrix theory approach to financial cross-correlations
        {\it Physica} A {\bf 287} 374 -- 382

\item[] Plerou V, Gopikrishnan P, Rosenow B, Amaral L A N, Guhr T and Stanley H E 2002
        A random matrix approach to cross-correlations in financial data
        \PR E {\bf 65} 066126-1 -- 066126-18

\item[] Santhanam M S and Patra P K 2001
        Statistics of atmospheric correlations
        \PR E {\bf 64} 016102-1 -- 016102-7

\item[] \v{S}eba P 2003
        Random matrix analysis of human EEG data
        \PRL {\bf 91} 198104-1 -- 198104-4

\item[] Siegel C L 1935
        Uber der analytische Theorie der quadratischen Formen
        {\it Ann. Math.} {\bf 36} 527 -- 606

\item[] Wishart J 1928
        The generalised product moment distribution in samples from a normal multivariate populations
        {\it Biometrika} A {\bf 20} 32 -- 52

\item[] Zabrodin A and Wiegmann P 2006
        Large-$N$ expansion for the $2D$ Dyson gas
        {\JPA} {\bf 39} 8933 -- 8963

\end{harvard}

\end{document}